\documentclass[fleqn,usenatbib]{mnras}

\usepackage{newtxtext,newtxmath}

\usepackage[T1]{fontenc}
\usepackage{ae,aecompl}

\usepackage{graphicx}	
\usepackage{amsmath}	
\usepackage{amssymb}	

\newcommand{\nuvr}{NUV-$r$}

\newcommand{\msun}{M$_{\odot}$}
\newcommand{\mstar}{$M_{\ast}$}
\newcommand{\lmstar}{log$\,$\mstar}

\newcommand{\dindex}{D$_n$(4000)} 
\newcommand{\ewhda}{EW(H$\delta_A$)}
\newcommand{\ewha}{EW(H$\alpha$)} 
\newcommand{\ha}{H$\alpha$}
\newcommand{\hb}{H$\beta$}
\newcommand{\hd}{H$\delta$} 
\newcommand{\Lha}{$L(\rm{H\alpha})$}

\newcommand{\rt}{$R_{\rm{t}}$}
\newcommand{\rhalf}{$R_{\rm{50}}$}
\newcommand{\rninety}{$R_{\rm{90}}$}
\newcommand{\re}{$R_{\rm{e}}$}
\newcommand{\rbar}{$R_{\rm{bar}}$}
\newcommand{\ba}{$b/a$}

\newcommand{\ebar}{$e_{\rm{bar}}$}

\newcommand{\hi}{H{\sc i}}
\newcommand{\htwo}{H$_{2}$}

\title[Central SF enhancement]{SDSS-IV MaNGA: the indispensable role of bars in enhancing the central star formation of low-$z$ galaxies}

\author[Lin et al.]{
Lin Lin$^{1}$\thanks{E-mail: linlin@shao.ac.cn}, 
Cheng Li$^{2}$\thanks{E-mail: cli2015@tsinghua.edu.cn}, 
Cheng Du$^{2}$, 
Enci Wang$^{3}$, 
Ting Xiao$^{4}$,
Martin Bureau$^{5,6}$,
\newauthor  
Amelia Fraser-McKelvie$^{7}$,
Karen Masters$^{8}$,
Lihwai Lin$^{9}$,
David Wake$^{10}$
and
Lei Hao$^{1}$
\\
$^{1}$ Shanghai Astronomical Observatory, Chinese Academy of Sciences, Shanghai 200030, China \\
$^{2}$ Department of Astronomy, Tsinghua University, Beijing 100084, China \\
$^{3}$ Department of Physics, ETH Zurich, Wolfgang-Pauli-strasse 27, CH-8093 Zurich, Switzerland \\
$^{4}$ Department of Physics, Zhejiang University, Hangzhou, Zhejiang 310027, China \\
$^{5}$ Sub-department of Astrophysics, University of Oxford, Denys Wilkinson Building, Keble Road, Oxford, OX1 3RH, UK \\
$^{6}$ Yonsei Frontier Lab and Department of Astronomy, Yonsei University, 50 Yonsei-ro, Seodaemun-gu, Seoul 03722, Republic of Korea \\
$^{7}$ School of Physics \& Astronomy, University of Nottingham, Nottingham NG7 2RD, U.K. \\
$^{8}$ Department of Physics and Astronomy, Haverford College, 370 Lancaster Ave, Haverford, PA 19041, U.S.A. \\
$^{9}$ Institute of Astronomy \& Astrophysics, Academia Sinica, Taipei 10617, Taiwan \\
$^{10}$ University of North Carolina at Asheville, Department of Physics,One University Heights, Asheville, NC 28804, USA
}

\date{Accepted XXX. Received YYY; in original form ZZZ}

\pubyear{2019}

\begin{document}
\label{firstpage}
\pagerange{\pageref{firstpage}--\pageref{lastpage}}
\maketitle

\begin{abstract}

 We analyse two-dimensional maps and radial profiles of \ewha, \ewhda, and \dindex\ of low-redshift galaxies using integral field spectroscopy from the MaNGA survey. Out of $\approx1400$ nearly face-on late-type galaxies with a redshift $z<0.05$, we identify 121 ``turnover'' galaxies that each have a central upturn in \ewha, \ewhda\ and/or a central drop in \dindex, indicative of ongoing/recent star formation. The turnover features are found mostly in galaxies with a stellar mass above $\sim$10$^{10}$ \msun\ and \nuvr\ colour less than $\approx5$. The majority of the turnover galaxies are barred, with a bar fraction of 89$\pm$3\%. Furthermore, for barred galaxies the radius of the central turnover region is found to tightly correlate with one third of the bar length. Comparing the observed and the inward extrapolated star formation rate surface density, we estimate that the central SFR have been enhanced by an order of magnitude. Conversely, only half of the barred galaxies in our sample have a central turnover feature, implying that the presence of a bar is not sufficient to lead to a central SF enhancement. We further examined the SF enhancement in paired galaxies, as well as the local environment, finding no relation. This implies that environment is not a driving factor for central SF enhancement in our sample. Our results reinforce both previous findings and theoretical expectation that galactic bars play a crucial role in the secular evolution of galaxies by driving gas inflow and enhancing the star formation and bulge growth in the center. 

\end{abstract}




\section{Introduction}


The bulges of disc-dominated galaxies are more complicated than previously thought. Initially believed to be similar to ellipitical galaxies, and thus formed via violent merging events and dominated by old stellar populations \citep[e.g.][]{Whitford-78, Renzini-99}, a large fraction of pseudobulges were later identified, built up from discs via bar- or spiral arm-driven processes \citep[e.g.][]{Peletier-Balcells-96, Helfer-03, Kormendy-Kennicutt-04}. Several studies then suggested that bulges can be composite systems in which a classical bulge and a pseudobulge co-exist, following a complex formation and evolutionary scenario \citep[e.g.][]{Athanassoula-05, Gadotti-09, Nowak-10, Kormendy-Barentine-10, Mendez-Abreu-14, Erwin-15}. Alternatively, direct observations of high-redshift galaxies suggest that bulges are formed through the inward migration and coalescence of clumps. These massive gas-rich clumps can sustain the rejuvenation of bulges, and act together with stellar migrations, minor mergers and other dynamical effects to finally build up bulges in the following several billion years \citep[e.g.][]{Noguchi-99, Carollo-07, Roskar-12, Bird-13, Mandelker-14, Mandelker-17}.

Tracing the central star formation of galaxies is key to understand how bulges build up their stellar masses. Any mechanism that causes disc instabilities and gas inflows may trigger central star formation, thus contributing to bulge growth. In the local Universe, the merger rate decreases as the Universe expands. Galaxy evolution is thus governed by long-term and secular processes. Internal mechanisms such as bars and spiral arms (and associated resonances) have been proposed to transfer angular momentum to the outer parts of discs, driving gas inflows toward the central regions \citep[e.g.][]{Athanassoula-03}. External processes like minor mergers can also cause disc instabilities that lead to gas flows towards the centres of the galaxies, eventually building up bulge components \citep[e.g.][]{Kormendy-Kennicutt-04}.

Many observations in the local Universe suggest that the existence of a bar-like structure in a galaxy contributes to cold gas inflows and the enhancement of central star formation \citep[e.g.][]{Sheth-05, Ellison-11, Wang-12}. Barred galaxies are indeed found to have higher molecular gas concentrations and systematically higher central star formation rates (SFR) than unbarred galaxies \citep[e.g.][]{Sakamoto-99, Sheth-05}.  \citet{Ellison-11} used Sloan Digital Sky Survey (SDSS) spectra to estimate the central SFRs of galaxies and compared them to global SFRs at given stellar mass, finding that barred galaxies have systematically higher central star formation than unbarred galaxies. \citet{Coelho-Gadotti-11} found that barred galaxies have a bimodal bulge age distribution, indicating an excess of young stars. 

Bars are however thought to suppress the global or galaxy scale star formation as well. Bar-induced star formation exhausts the infalling gas quickly, helping to build-up a central bulge, that can in turn stabilise the disc and stop further gas falling in \citep[e,g.][]{Bournaud-Combes-02, Athanassoula-03}. Observations show that a bar-like structures are preferentially found in massive, gas-poor galaxies rather than in gas-rich galaxies \citep[e.g.][]{Masters-12, Cheung-13, CervantesSodi-17}. \citet{Wang-12} thus found that galaxies with strong bars can either enhance central star formation or suppress it, and the stellar bars have been proposed to play an important role in star formation quenching of local isolated galaxies \citep[e.g.][]{Masters-11, Gavazzi-15, George-19, Newnham-20}.

Despite the above, the connection between bars and central nuclear activity is still unclear. Numerous studies suggest that bars are not related to active galactic nucleus (AGN) \citep[e.g.][]{Cisternas-15, Cheung-15}, while others claim there is a weak connection \citep[e.g.][]{Knapen-00, Laine-02, Galloway-15}. Nuclear, or inner bars would be needed to take material from the galactic scale bar to the supermassive black hole scale, which could explain this mismatch.

Apart from bar-induced central star formation, galaxy interactions/disruptions are another well-known mechanism for enhancing star formation, as strong starbursts are predominately found in merging systems. Nevertheless, several studies suggest that enhanced star formation rates may not be ubiquitous in interacting galaxies. The average SFR enhancement in pair galaxies is below a factor of 2-3 \citep[e.g.][]{Lin-07, Li-08, Ellison-08, Knapen-15}. Many factors, such as the separation of the companions, gas fraction, mass ratio of the systems or even the merging geometry affect the star formation enhancement \citep[e.g.][]{Ellison-08, Davies-15, Pan-18}. Projected separation of less than $\sim$100 kpc is required for central enhancement \citep{Li-08}. 
    
The recent development in the field of integral-field spectrographs (IFSs) provides us with a unique opportunity to revisit the detailed morphological substructures in the inner regions of galaxies, including bulges and bars. In a previous work \citep{Lin-17}, we used the 4000\AA\ break (\dindex), the equivalent width of \hd\ absorption (\ewhda) and \ha\ emission (\ewhda) to reveal the recent star formation history of 57 galaxies selected from Calar Alto Legacy Integral Field Area Survey \citep[CALIFA;][]{Sanchez-12}. By analysing their two-dimensional maps and radial profiles, we identified a class of ``turnover'' galaxies that each have a significant inner drop in \dindex\ and/or a corresponding upturn in \ewha\ and \ewhda. This indicates that the central regions of these galaxies have experienced star formation in the past 1-2 Gyr. We found most turnover galaxies to host a bar structure, but only half of barred galaxies to have a central turnover. In the following work of \citet{Chown-19}, we studied the link between star formation enhancements and molecular gas surface densities. Turnover galaxies are associated with higher molecular gas concentrations, supporting bar- or interaction-driven gas inflow scenarios. However, due to the limited sample size in that study, turnover features in recent star formation indicators are still not fully understood. It is therefore necessary to extend previous work with a larger sample covering a wider parameter space and enabling us to build a control sample. 

In this paper, we use the Mapping Nearby Galaxies at Apache Point Observatory \citep[MaNGA; ][]{Bundy-15} data set to accomplish this, and confirm that there is a strong link between turnover features and the bars. This paper is organised as follows. Section~\ref{sec:data} describes our data and sample selection, as well as our measurements of bar properties, the environment catalogue we adopted, and how we identify turnover features. We present the main result in Section~\ref{sec:result}. The discussion and conclusions are addressed in Section~\ref{sec:discuss} and \ref{sec:conclusion}, respectively. Throughout this paper, we adopt a flat $\Lambda$ cold dark matter cosmology, with parameters $\Omega$ = 0.3, $\Lambda$ = 0.7 and $H_{0}$ = 70 km s$^{-1}$ Mpc$^{-1}$.

\section{Data and Sample Selection}
\label{sec:data}

\subsection{Overview of MaNGA}

As one of the three key programs of the fourth-generation of SDSS \citep[SDSS-IV;][]{Blanton-17}, MaNGA is an integral-field unit (IFU) spectroscopic survey \citep{Bundy-15}, aiming to achieve kpc-scale spatially resolved IFU datacubes for about 10,000 nearby galaxies. The MaNGA instrument has 29 IFUs including 12 seven-fibre mini-bundles for flux calibration and 17 science bundles with multiple fields of view ranging from 12\arcsec\ to 32\arcsec\ on sky \citep{Drory-15}. The spectra  are obtained using the two dual-channel BOSS spectrographs at the 2.5-m Sloan Telescope \citep{Gunn-06, Smee-13}, covering wavelengths from 3622 to 10,354 \AA\ with a spectral resolution of $R\approx2000$, and reaching a target $r$-band signal-to-noise ratio SNR $= 4-8$ (per \AA\ per 2\arcsec-fibre) at 1-2 effective radius (\re) with a typical exposure time of three hours.

MaNGA target galaxies are selected from NASA-Sloan Atlas catalog\footnote{http://www.nsatlas.org} \citep[NSA;][]{Blanton-11}, which is
a SDSS-based galaxy catalogue including physical parameters of 640,000 galaxies from GALEX, SDSS and 2MASS. The target sample is well defined to have a statistically significant size, a uniform spatial coverage in units of \re,  and an approximately flat total stellar mass distribution between 10$^{9}$ and 10$^{11}$ \msun. Details of the sample design and optimization can be found in \citet{Wake-17}. The MaNGA sample includes three subsamples. The Primary and Secondary samples are selected to have a flat distribution in $i$-band absolute magnitude, and a spatial coverage out to 1.5 and 2.5 \re, respectively. The third Color Enhanced sample increases the fraction of rare populations in the galaxy colour-magnitude diagram, such as high-mass blue galaxies and low-mass red galaxies. The median redshift of the MaNGA galaxies is $z\approx0.03$.

The MaNGA observing strategy, survey execution and flux calibration are described in \citet{Yan-16a, Yan-16b}. The absolute flux calibration is better than 5\% for more than 80\% of the wavelength range.  Raw data are reduced with the Data Reduction Pipeline \citep[DRP;][]{Law-16}, that produces a datacube for each galaxy with a spaxel size of 0\farcs5 and an effective spatial resolution of $\approx$ 2\farcs5. The MaNGA team has developed a Data Analysis Pipeline (DAP) that performs full spectral fitting to the spectra in DRP datacubes and calculates a variety of physical parameters including stellar kinematics, emission-line properties and spectral line indices \citep{Westfall-19, Belfiore-19}.  

In this work we use the internal data release of MaNGA, the MaNGA Product Launch 7 (MPL-7), that includes DRP and DAP products for
4672 galaxies and is identical to the MaNGA data released with the SDSS Data Release 15 \citep[DR15;][]{Aguado-19}.

\subsection{Sample selection}

\begin{figure}
	\begin{center}
		\includegraphics[width=0.45\textwidth]{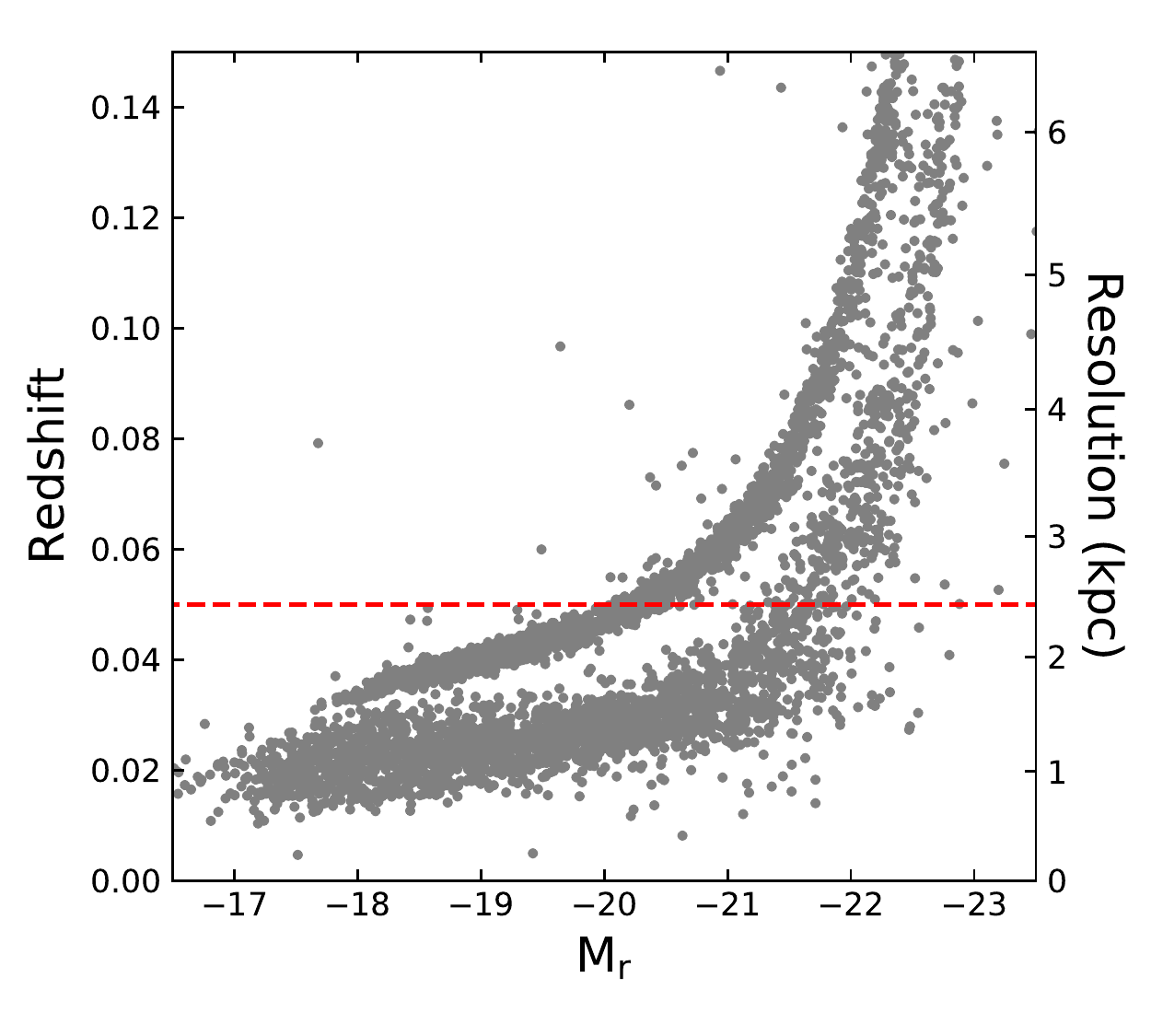}
	\end{center}
	\caption{Distribution of MPL-7 galaxies on the diagram of redshift vs. $r$-band absolute magnitude. The spatial resolution, corresponding to the physical scale of a typical PSF FWHM of 2\farcs5, is also shown along the ordinate. The galaxies populate two branches, corresponding to the MaNGA Primary and Secondary samples. To ensure sufficient spatial resolution, we limit our sample galaxies at redshifts $z<$ 0.05 with axial ratios \ba\ $>$ 0.5.}
	\label{fig:resolution}    
\end{figure}

\begin{table*}
	\centering
    \caption{Summary of multiple samples used in this paper.}
    \label{tbl:numbers}
    \begin{tabular}{llc}
        \hline
        Sample &  Type  &  Number \\
        \hline
        MPL-7  		& $z$$<$0.05, $b/a$$>$0.5, T-type$>$0 			&  1425	 \\
  					& barred/unbarred	  				&  252/1173  \\
  					& pair stage 1/2/3      			&  93/22/113 \\
  					& SF/comp/LINER/Sy/--$^\mathrm{a}$ 				&  902/145/108/74/196  \\
        \hline
        Turnover galaxies	& All			 			&  121   \\
					&  \ewha/\ewhda/\dindex-detected	&  109/74/101   \\
  					& barred/unbarred					&  108/13   \\
					& pair stage 1/2/3                	&  5/5/10   \\
  					& SF/comp/LINER/Sy/--				&  30/19/39/25/8   \\ 
        \hline					
        Control sample	& \nuvr\ and \mstar\ matched     &  121  \\
  					& barred/unbarred					&  41/80   \\
					& pair stage 1/2/3                	&  9/2/13   \\
					& SF/comp/LINER/Sy/-- 				&  36/18/36/16/15   \\
		\hline
		\multicolumn{3}{l}{(a) `--' means lack of emission lines for BPT classification.}  
	\end{tabular}
\end{table*}

In this work we will focus on the central regions of galaxies, thus requiring a statistical sample of data  with sufficient spatial resolution. Figure~\ref{fig:resolution} shows the distribution of the MPL-7 galaxies on the diagram of redshift vs. $r$-band absolute magnitude. The spatial resolution in kpc is plotted on the ordinate, corresponding to the 2\farcs5 full width at half maximum (FWHM) of the MaNGA effective point spread function (PSF). The galaxies are located in two separate sequences in the figure, corresponding to MaNGA Primary and Secondary samples described above. 

For our study, we select a sample of galaxies from the MPL-7 as follows. First, we select nearly face-on galaxies, with minor-to-major axial ratio \ba\ $>$ 0.5, where the semi-major ($a$) and semi-minor ($b$) axes are taken from the NSA and are measured at the 25 mag arcsec$^{-2}$ isophote in the $r$-band. Second, we restrict ourselves to relatively low redshifts $z$ $<$ 0.05. As can be seen from Figure~\ref{fig:resolution}, this redshift cut corresponds to a worst spatial resolution of $\approx$2.3 kpc, comparable to the typical size of galactic bulges and bars \citep{Gadotti-11}. This choice of the upper limit on redshift is a trade-off between having a substantially large sample for good statistics and ensuring a relatively high spatial resolution for measuring central  structures. More discussion on spatial resolution can be found in Appendix~\ref{app:resolution}. Third, we select late-type galaxies by requiring T-type $>$ 0. The morphological classification was taken from \citet{DomnguezSanchez-18}.  Finally, we make use of the MPL-7 version of the close pair galaxy catalogue (see Section~\ref{sec:data:pair} for details) to exclude 78 galaxies that are merging systems at the final coalescence stage. These galaxies do not have regular morphologies/structures/centres, and it is not straightforward to define radial profiles and make comparisons to other types of galaxies.

These restrictions give rise to a sample of 1452 late-type galaxies, which forms the main sample to be studied below. The numbers of galaxies in different categories are listed in Table~\ref{tbl:numbers}. In this work, we use data produces provided by the MaNGA DAP.  Specifically, we use the ``hybrid'' binning (HYB10) products, for which the stellar spectrum (continuum plus absorption lines) and stellar kinematics are obtained by applying the Penalized PiXel-Fitting (pPXF) code \citep{Cappellari-17} to spectra binned to S/N$\sim$10.  For each spaxel in the datacube, the best-fitting stellar spectrum is  subtracted from the observed spectrum and the starlight-subtracted spectrum is then used to measure emission lines by fitting each line with a Gaussian profile \citep{Belfiore-19}. We remove spaxels that might be contaminated by foreground stars or have critical failures during data reduction, using the same bitmask as adopted in the DAP. 

Based on the \ewha, \ewhda\ and \dindex\ maps from the MaNGA DAP, we estimate the one-dimensional radial profiles of these quantities by adopting a constant spatial sampling of 0\farcs5 along the major axis, with fixed ellipticity and position angle and thus perfectly concentric elliptical annuli for all radial bins. The global ellipticity, position angle of the major axis and the major-to-minor axis ratio are taken from the NSA catalogue. Uncertainties on the derived quantities are given by  the 1$\sigma$ scatter of the spaxels within each radial annulus. We note that the one-dimensional profile is mixing non-axisymmetric structures, so it is important to examine the two-dimensional map simultaneously.

\subsection{Bar classification and measurement}
\label{sec:data:bar}

\begin{table}
	\centering
	\caption{Measurements of barred galaxies. }
	\label{tbl:bar}
	\begin{tabular}{crrr}
		\hline
		ID & $r_{\rm bar}$ & $e_{\rm bar}$ & PA$_{\rm bar}$ \\
		 & (\arcsec) & & ($^{\circ}$) \\
		 (1) & (2) & (3) & (4) \\
		\hline
		9487-12701 & 5.9$\pm$0.6 & 0.50$\pm$0.05 & 114.9$\pm$2.3 \\
		8134-12701 & 4.8$\pm$0.5 & 0.60$\pm$0.06 & 83.1$\pm$1.1 \\
		8728-12701 & 8.6$\pm$0.9 & 0.57$\pm$0.06 & 113.4$\pm$1.0 \\
		8338-12701 & 7.1$\pm$0.7 & 0.62$\pm$0.06 & 120.4$\pm$0.9 \\
		9888-12701 & 12.6$\pm$1.2 & 0.61$\pm$0.06 & 60.1$\pm$0.4 \\
		 ...       &  ...  & ...  & ... \\
		8713-9102 & 7.8$\pm$0.8 & 0.50$\pm$0.05 & 5.4$\pm$0.8 \\
		8486-9102 & 2.26$\pm$0.23 & 0.50$\pm$0.05 & 77.04$\pm$1.78 \\
		8442-9102 & 7.80$\pm$0.78 & 0.58$\pm$0.06 & 86.17$\pm$0.96 \\
		8315-9102 & 4.84$\pm$0.48 & 0.60$\pm$0.06 & 27.14$\pm$0.70 \\
		9487-9102 & 8.6$\pm$0.8 & 0.64$\pm$0.06 & 53.0$\pm$1.0 \\

	\hline 
	\multicolumn{4}{l}{Note: Full table is available online in machine-readable format.} \\
	\multicolumn{4}{l}{(1) MaNGA ID. } \\  
	\multicolumn{4}{l}{(2) bar length and uncertainty. } \\ 
	\multicolumn{4}{l}{(3) bar ellipse and uncertainty. } \\ 
	\multicolumn{4}{l}{(4) bar position angle and uncertainty. } \\ 
	\end{tabular}
\end{table}

We identify galactic bars in our sample and measure their lengths and ellipticities by applying the commonly-used Image Reduction and Analysis Facility (IRAF) task {\tt ELLIPSE} to the background-subtracted $r$-band SDSS images. The {\tt ELLIPSE} task fits two-dimensional ellipses to the isophotes for each galaxy elliptical annuli, at logarithmically-spaced radii along a galaxy major axis. For each galaxy in our sample, we adopt the sampling radius with a step of 1.1, and the centre of each ellipse is allow to vary. We apply the {\tt ELLIPSE} task twice, adopting \rhalf\ and \rninety, which taken from the NSA catalogue, as the starting radius respectively. \rhalf\ and \rninety\ are the elliptical Petrosian radii enclosing 50 and 90 per cent of the total light in $r$-band. Then the two profiles are merged in order to fully cover the radial profile. Following \citet{Lin-17}, we identify the bar structure according to the presence of an abrupt decrease of the ellipticity profile and an associated change of position angle (P.A.) --- a galaxy is classified as having a bar if the ellipticity increases above 0.25 as the radius increase from the galactic centre outwards, before decreasing by at least 0.1 at some radius. Meanwhile, the P.A. is required to change more than 10$^{\circ}$ from the end of the bar to the outer disc region.

We visually examined the SDSS $gri$ images of all 1425 galaxies in our sample. As noted by \citet{Barazza-Jogee-Marinova-08} and \citet{Wang-12}, some barred galaxies (though few in number) may be missed by the ellipse-fitting procedure, probably due to the existence of a bulge or the alignment of the bar with the minor axis at a moderate inclination angle (making the bar appear rounder and smoother). In these cases, the ellipticity would be underestimated and so would not fulfill the requirement of ellipticity $>$ 0.25. Galaxies in which the bar smoothly connects with the spiral arms or ring would also be missed.  After visually examining all the images, we include 15 additional galaxies with a weak bar, 6 per cent of our final barred galaxy sample. 

Our final barred galaxy sample consists of 252 galaxies.  The half-length ($r_{\rm bar}$ in arcseconds), ellipticity ($e_{\rm bar}$) and P.A. (PA$_{\rm bar}$ in degrees) of each bar is determined from the best-fitting ellipse at the semi-major axis radius at which the ellipticity reaches a local maximum\footnote{The bar length determined by the location of the maximum ellipcity is thought to be a lower limit \citep{Erwin-05}. However, it is well correlated with the true value. The correlation we find in the following should not be affected. }. We further calculate the deprojected bar length and ellipticity following \citet{Gadotti-07}. We adopt uncertainties of 10\% in both the bar length and ellipticity measurements, taking into account the uncertainties due to different bar measurement methods \citep{Zou-14}.  Our bar measurements are listed in Table~\ref{tbl:bar}.

The bar fraction in the main sample is 18\% (252/1425), which seems much lower than the known fraction of 1/3 for strong bars or the fraction of 2/3 for all bars. We note that the bar fraction is a strong function of stellar mass and sSFR. After the selection criteria, our main sample is restricted to late-type galaxies and have a median stellar mass of $\sim$10$^{10}$\msun. The bar fraction in this type of galaxies is $\sim$20\% in \cite{Cheung-13}, close to our bar fraction. In addition, our bar fraction in control sample is 34\% (41/121). The increase in the bar fraction can be understood from the larger stellar mass in control sample. Overall, our bar fraction is consistent with the results from more general population.

\subsection{Environment catalogues}
\label{sec:data:pair}

To investigate potential environmental effects on our galaxies at different physical scales, we make use of the MaNGA galaxy pair catalogue of \citet{Pan-19}, the SDSS galaxy group catalogue of \citet{Yang-07} and the local environment density measure inferred from the density field reconstructed by \citet{Wang-16}. We briefly describe these catalogues below.

The close galaxy pair catalogue was constructed using the MPL-7 (or SDSS DR15) version of the MaNGA sample as described in detail in \citet{Pan-19}.  A galaxy pair is defined as a system of two galaxies with a projected separation less than 50 kpc $h^{-1}$ and a line-of-sight velocity difference less than 500 km s$^{-1}$. The pairs are further classified into four different {\it merger stages} based on a visual inspection of their SDSS images.  Galaxies in pairs at Stage 1 are those without any morphology distortion. Galaxies in pairs at Stage 2 have obvious tidal tails or bridges. Galaxies in pairs at Stage 3 have only weak morphology distortions.  Stage 4 encompasses galaxy mergers close to the final coalescence stage, for which the two galaxies are largely overlapping, meaning it is hard to measure radial profiles. Therefore, we exclude Stage 4 in this study and focus on the well-separated pairs at Stages 1-3.

The galaxy group catalogue of \citet{Yang-07} has been widely used in previous studies. Galaxy groups are identified by applying a modified version of the halo-based group-finding algorithm of \citet{Yang-05} to the galaxy sample of SDSS Data Release 7 with 0.01 $<z<$ 0.2, with a redshift completeness greater than 70\%.  Following common practice, for each group we take the most massive galaxy as the central galaxy and consider the other member galaxies as satellites. Stellar masses are taken from the NSA catalog \citep{Blanton-11}.

We also use the three-dimensional reconstructed local mass density ($\delta=\rho/\bar{\rho}$, where $\rho$ is the local matter density and $\bar{\rho}$ is the mean matter density) to quantify the density of the local environment for each galaxy in our sample. The mass density field is taken from the Exploring the Local Universe with the reConstructed Initial Density Field project \citep[ELUCID;][]{Wang-16}, that by construction can very well reproduce the observed distributions of both galaxies and groups of galaxies. ELUCID provides local densities measured at different scales. In this work, we use the density estimated by smoothing the density field with a Gaussian kernel at a scale of 1 Mpc $h^{-1}$.

\section{Results}
\label{sec:result}

\subsection{Identification of ``turnover'' galaxies}

\begin{figure*}
	\begin{center}
		\includegraphics[width=\textwidth]{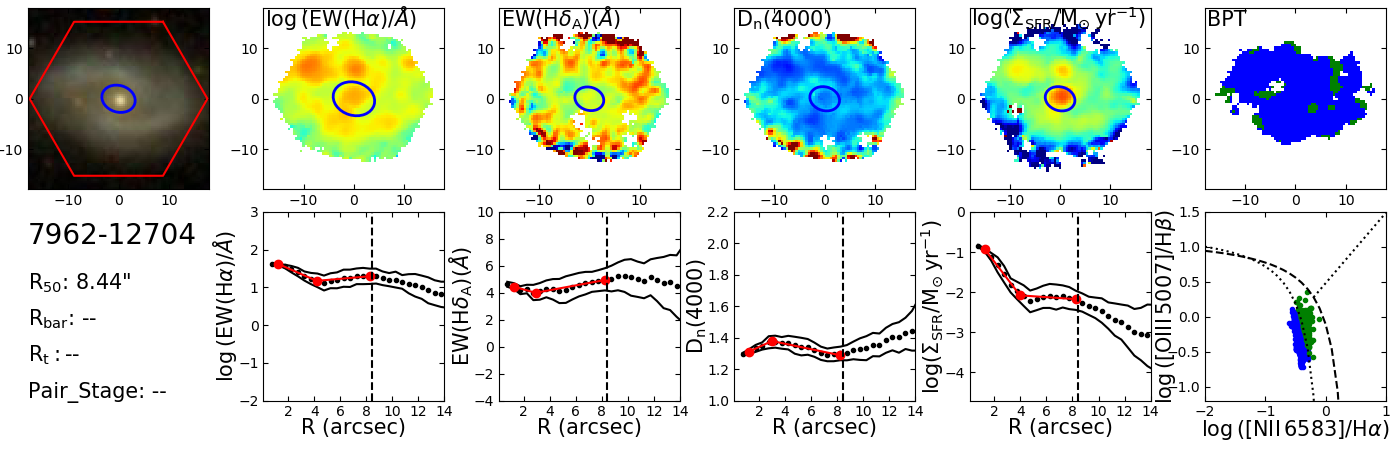}
		\includegraphics[width=\textwidth]{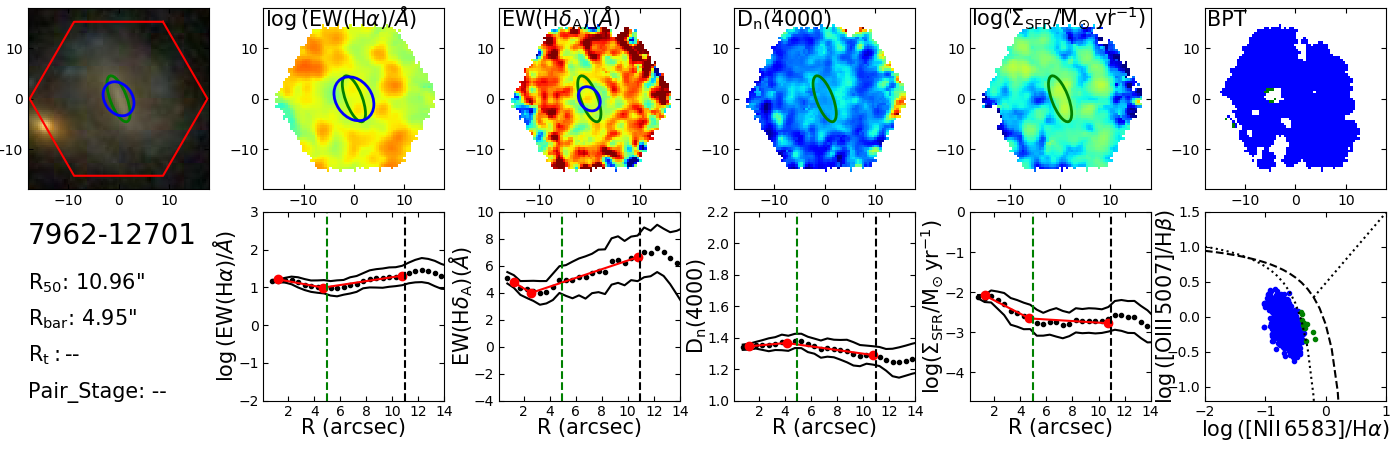}
		\includegraphics[width=\textwidth]{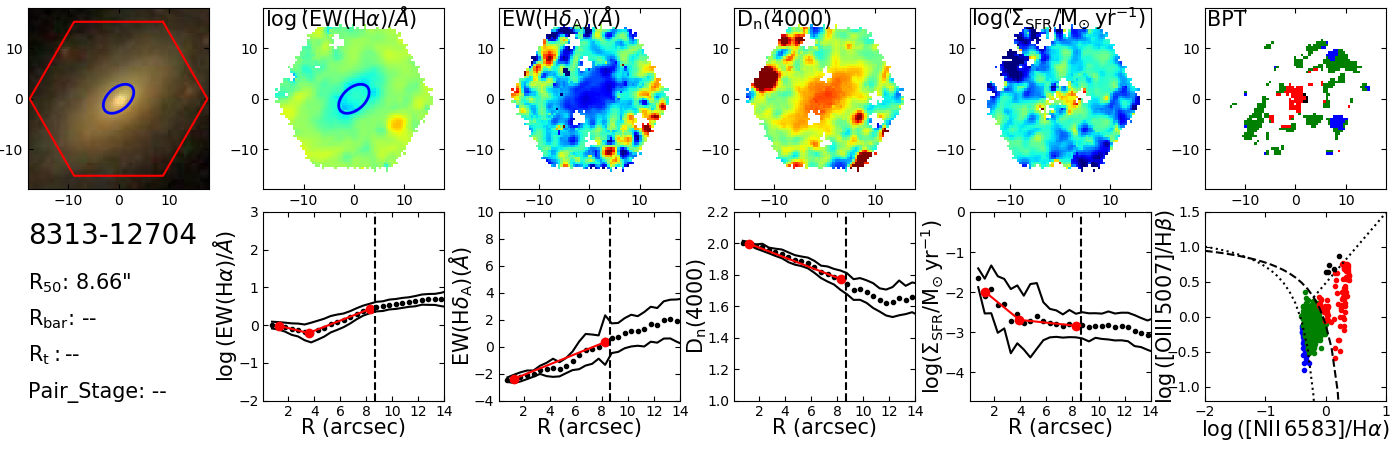}
	\end{center}
	\caption{Three examples of turnover galaxy candidates excluded after visual inspection. From left to right: Optical $gri$ image, maps and profiles of \ewha, \ewhda, \dindex\ and SFR surface density profile (measured from dust-corrected H$\alpha$ emission), and resolved BPT diagram. The red hexagons correspond to the field of view of the MaNGA bundle. In each map the blue and green ellipses indicate respectively the potential turnover radii and (if present) the bar. In each radial profile, the solid and dotted black lines are the median value and 1$\sigma$ scatter in each annulus, the segment fitting is plotted as solid red lines, the green (if present) and black vertical dashed lines indicate respectively the end of the bar (\rbar) and the effective radius (\rhalf). The first galaxy has a break radius in all the three indicators, but it happens at the bulge-disc transition (by visual inspection). We thus conclude it is not a genuine turnover within bulge. The second galaxy has break radius within \rhalf, but the profiles are straight within \rbar. The last galaxy is excluded because it lacks of a robust feature. More details are provided in the text. }
	\label{fig:excluded}
\end{figure*}

We firstly identify ``turnover'' galaxies in our sample that present a significant star formation enhancement in their centres, applying a selection procedure similar to that in our previous work \citep[][hereafter Paper I]{Lin-17}. In short, the identification procedure comprises two steps. In the first step, we select {\it potential} turnover galaxies, that each show a central upturn or drop in the one-dimensional radial profiles of three parameters \ewha, \ewhda\ and \dindex. Next, we visually examine the SDSS $gri$-image, two-dimensional maps, and the radial profiles of each galaxy, and decide whether or not the turnover feature indeed occurs in the inner regions. Here, ``inner regions'' mean the inner parts of the disc, including both the bulge and bar-dominated regions (if present). In the rest of this subsection, we describe the identification process in more detail.

For a given radial profile, we start from the outer most region by fitting a linear function to the three data points at the largest radii. We then determine whether the next data point at the neighbouring smaller radius follows the same trend, by evaluating the deviation of that data point from the best-fitting line. If this deviation is less than the 1$\sigma$ scatter of all the spaxels at the same radius as the data point, we perform the linear fit again but now include the new data point. In this way, the data points at smaller radii are gradually added to the fit, one by one. If there is a newly added data point deviated significantly from the best-fitting line by more than 1$\sigma$, indicating a sudden break in the slope of the profile. From this radius, we repeat the above process, start fitting a new line with a different slope to the next three data points and gradually adding more data points at smaller radii. We thus consider the whole profile from the outside in by adding data points one-by-one with a slightly adjusted slope or an entirely new slope until the radius reaches half of the PSF FWHM (1\farcs25). 

This procedure was applied to the profiles of \ewha, \ewhda\ and \dindex\ for every galaxy in our sample.  We then select turnover galaxy candidates  by requiring at least one of the three parameters to have a slope break in its profile (increment for \ewha\ and \ewhda, and turn-down for \dindex\ as the radius decreases) within \rhalf\ or \rbar\ if \rbar>\rhalf. This restriction to the inner regions is based on our empirical finding from Paper I that nearly all turnover radii are smaller than \rhalf\ or \rbar. We thereby select 223 turnover galaxy candidates out of the 1425 galaxies in our sample.

We then visually inspect the SDSS $gri$-image of each turnover galaxy candidate, overplotting the isophotal ellipse at the break radius onto the image. Much attention is paid to the location of the break relative to the bulge, bar and disc components. A candidate is selected as a {\it real} turnover galaxy if the inner most break happens within the bulge radius if the galaxy is unbarred, or within the bar length if it is a barred galaxy.  This visual inspection was carried out independently by two of the authors (LL and CD). In this step, 102 candidates are excluded for various reasons, as illustrated in Figure~\ref{fig:excluded} showing for three example galaxies the optical image, the two-dimensional maps of \ewha, \ewhda, \dindex, log$(\Sigma_{\rm SFR})$ and Baldwin, Phillips \& Telervich diagrams \citep[BPT;][]{Baldwin-81} , and the corresponding radial profiles. The SFRs are calculated from dust-corrected H$\alpha$ fluxes (see \autoref{sec:enhance_sf} for details). In the majority of these galaxies (72/102), the slope break of the radial profile reflects the transition between bulge and disc, rather than a central turnover feature within the bulge. This is the case for the first galaxy (MaNGA ID 7962-12704) in Figure~\ref{fig:excluded}. This galaxy shows a break at $\approx$4$\arcsec$ in \ewha, \ewhda\ and \dindex\ at a radius that  that is much larger than the central bright core, which is likely a pseudo-bulge with size of $\sim$1$\arcsec$ but shows no obvious turnover features in the profiles. There are 24 candidates are even bulgeless, thus they are also excluded. The breaks are due to fluctuations among spiral arms. For example, the second galaxy (MaNGA ID 7962-12701) in Figure~\ref{fig:excluded} is a bulgeless barred galaxy. It has a break within \rhalf, indicated by the vertical black dashed line, but the profiles within the bar (indicated by the green dotted line) are roughly flat, with no obvious turnover feature.  Finally, a few galaxies (6/102) are excluded because this slope changes are too gentle and smooth, and are thus unlikely caused by significant changes of stellar populations. As shown in the bottom panels, the selected galaxy (MaNGA ID 8313-12704) does have  a variation of the slope of its \ewha\ profile, but the signal is no obvious.

\begin{figure}
	\begin{center}
		\includegraphics[width=0.45\textwidth]{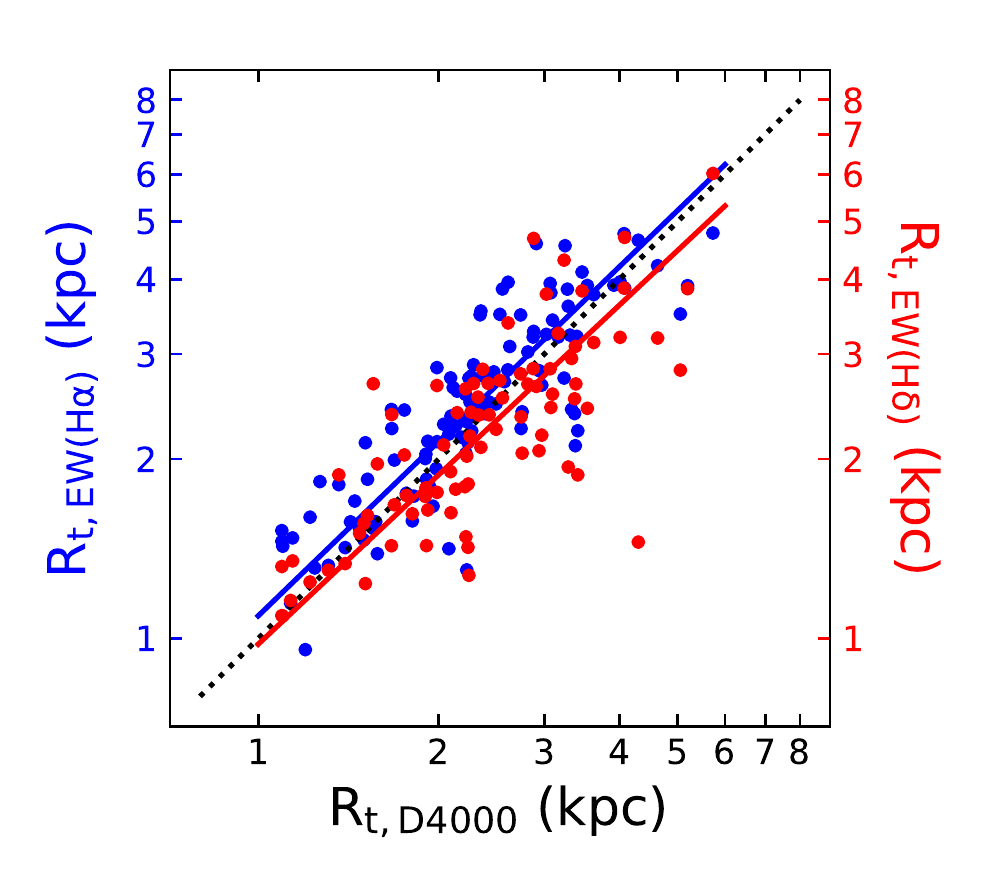}
		\caption{Comparison between the different turnover radii of our final turnover sample galaxies. The black dotted line is the one-to-one line, the blue and red solid lines are the linear fittings between $R_{\rm t, D4000}$ and $R_{\rm t, EW(H\alpha)}$, $R_{\rm t, D4000}$ and  $R_{\rm t, EW(H\delta)}$, respectively. Different measurements are highly consistent with each other, with a typical 1$\sigma$ scatter of 0.11 dex. }
		\label{fig:Rt_diff_indicators}
		
	\end{center}
\end{figure}

\begin{figure*}
	\begin{center}
		\includegraphics[width=\textwidth]{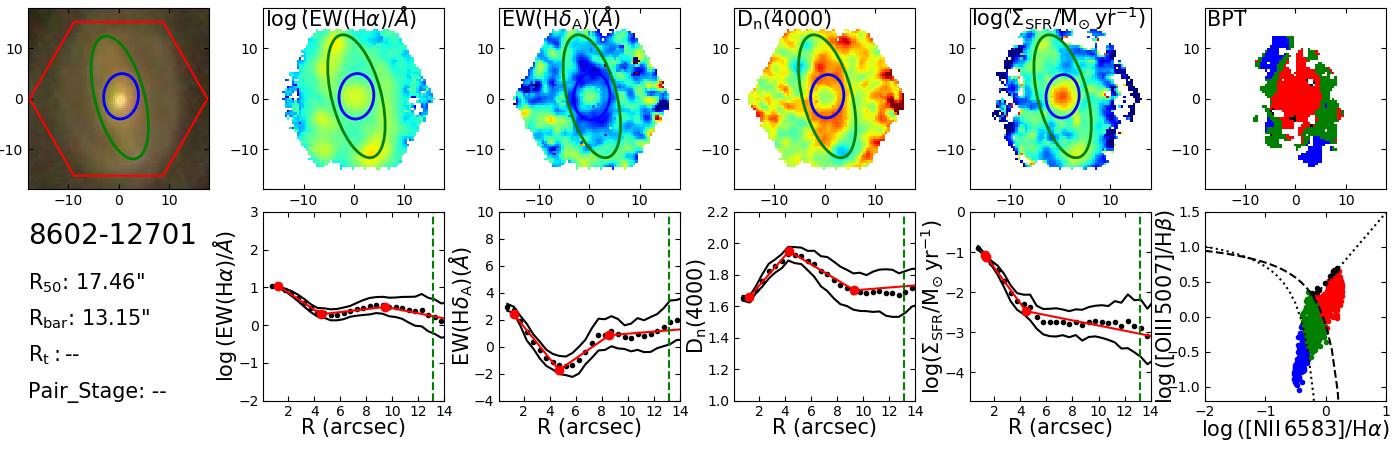}
		\includegraphics[width=\textwidth]{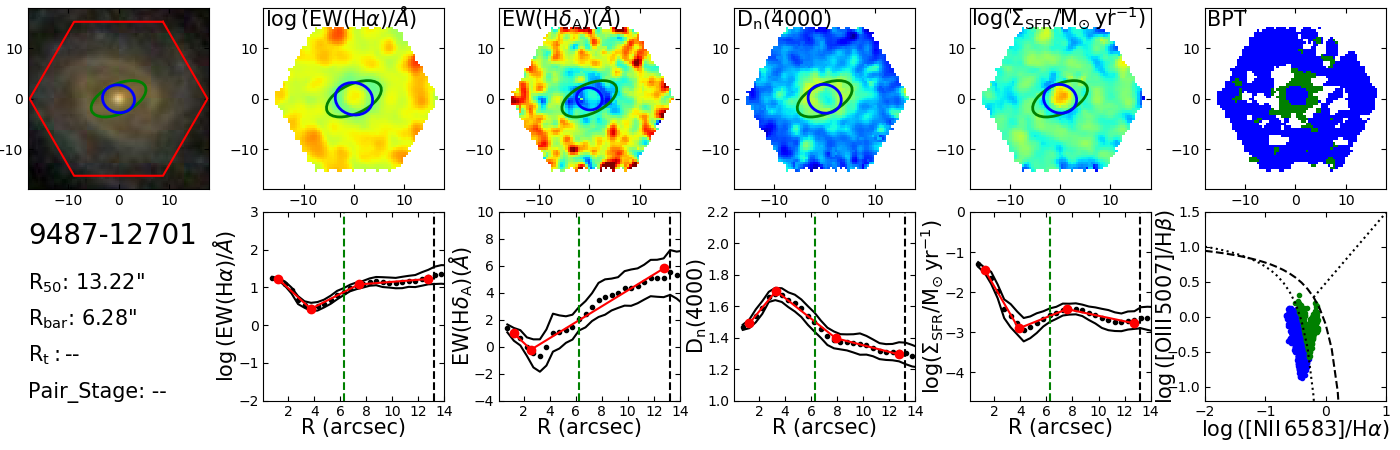}
		\includegraphics[width=\textwidth]{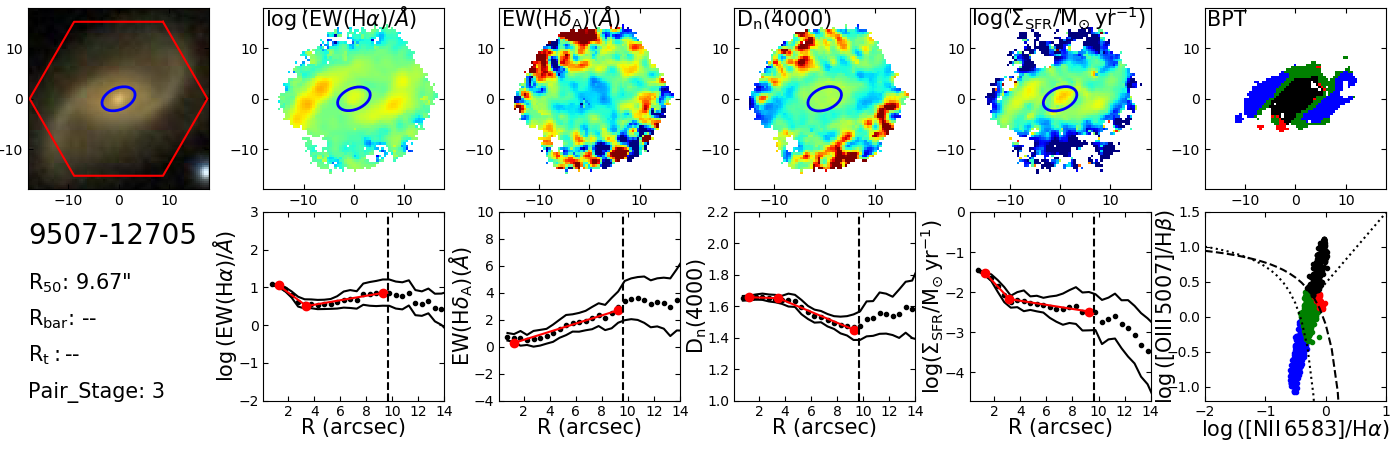}
		\includegraphics[width=\textwidth]{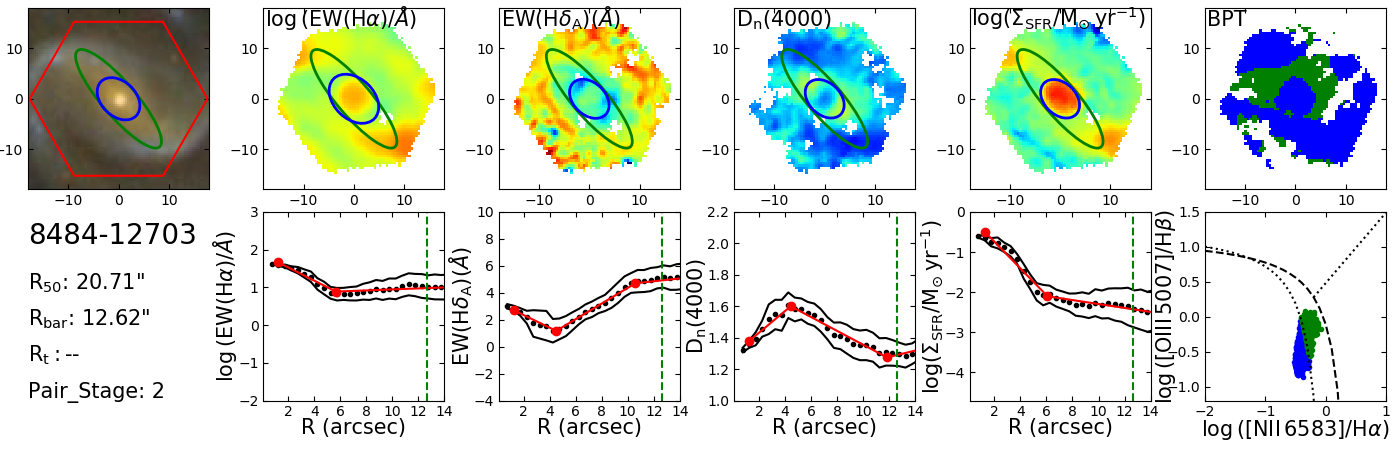}
	\end{center}
	\caption{Four examples of turnover galaxies. From left to right: Optical $gri$ image, maps and profiles of \ewha, \ewhda, \dindex\ and SFR surface density (measured from dust-corrected H$\alpha$ emission), and resolved BPT diagram. Labels and symbols are the same as Figure~\ref{fig:excluded}. The first two are barred galaxies, which are selected to demonstrate central SF and LINER-emission. The last two are disturbed galaxies in pair systems. The central SFRs could be contaminated by nuclear activity if they have LINER or Syfert centres.}
	\label{fig:2d_maps}    
\end{figure*}

\begin{table*}
	\centering
	\caption{Turnover property measurements.}
	\label{tbl:turnover}
	\begin{tabular}{cccccc} 
		\hline
		ID & R$_{\rm t}$ & $\Delta$$\,$log$\,$EW(H$\alpha$) &  $\Delta$$\,$EW(H$\delta$) & $\Delta$$\,$D$_{\rm n}$4000 &  $\Delta$$\,$log$\,$SFR$_{\rm <R_{\rm t}}$ \\
		& (\arcsec) & (\AA) & (\AA) & & (M$_{\odot}$ yr$^{-1}$) \\
		(1) & (2) & (3) & (4) & (5) & (6) \\
		\hline
		9487-12701&     3.2$\pm$0.5&     1.2$\pm$0.1&     2.0$\pm$1.2&    -0.34$\pm$0.07&     1.0$\pm$0.3	\\
		8728-12701&     3.3$\pm$0.4&     0.7$\pm$0.2&     1.2$\pm$0.8&    -0.12$\pm$0.05&     0.8$\pm$0.3	\\
		8155-12701&     7.3$\pm$2.0&     1.2$\pm$0.4&     0.9$\pm$1.1&    -0.21$\pm$0.12&     0.7$\pm$0.5	\\
		8715-12701&     5.0$\pm$0.4&     0.6$\pm$0.2&     2.0$\pm$1.1&    -0.15$\pm$0.08&     1.2$\pm$0.5	\\
		8952-12701&     3.1$\pm$0.1&     0.1$\pm$0.1&     1.7$\pm$0.5&    -0.13$\pm$0.03&      ... 	\\
		     ...  &     ...  &     ...  &     ...  &     ...  &      ... 	\\
		9869-9102&     7.0$\pm$0.1&     1.2$\pm$0.3&    -0.3$\pm$0.8&    -0.48$\pm$0.10&     1.6$\pm$0.5	\\
		8442-9102&     4.0$\pm$0.9&     1.0$\pm$0.2&     1.4$\pm$0.9&    -0.09$\pm$0.04&     0.5$\pm$0.3	\\
		8713-9102&     3.9$\pm$0.5&     1.4$\pm$0.1&     3.8$\pm$1.0&    -0.39$\pm$0.10&     1.1$\pm$0.2	\\
		8444-9102&     3.4$\pm$0.2&     0.5$\pm$0.2&     0.2$\pm$0.8&    -0.08$\pm$0.05&     0.9$\pm$0.4\\
		9487-9102&     3.6$\pm$0.8&     0.6$\pm$0.3&     1.7$\pm$0.7&    -0.26$\pm$0.09&     0.7$\pm$0.4	\\
		\hline 
		\multicolumn{4}{l}{Note: Full table is available online in the machine-readable format.} \\
		\multicolumn{4}{l}{(1) MaNGA ID. } \\  
		\multicolumn{4}{l}{(2) turnover radius and uncertainty. } \\  
		\multicolumn{4}{l}{(3) log EW(H$\alpha$) enhancement and uncertainty. } \\ 
		\multicolumn{4}{l}{(4) EW(H$\delta)$ enhancement and uncertainty. } \\ 
		\multicolumn{4}{l}{(5) D$_{\rm n}$(4000) drop and uncertainty. } \\ 
		\multicolumn{4}{l}{(6) log SFR enhancement within turnover radius and uncertainty. } \\ 
	\end{tabular}
\end{table*}

Our final sample includes 121 galaxies selected following the above two-step scheme based on \ewha, \ewhda\ and \dindex\ independently, and each presenting significant central turnover feature in at least one of the three parameters. However, we find that in most cases the turnover feature is simultaneously present in all the three parameters. The number of turnover galaxies identified from each parameter is listed in Table~\ref{tbl:numbers}.  More than 80\% are identified from at least two parameters. In addition, the turnover radii measured from these different parameters are highly consistent with each other. This can be seen from Figure~\ref{fig:Rt_diff_indicators}, comparing the turnover radii determined from the three different parameters. The typical scatter is 0.11 dex. In what follows,  if one galaxy has different turnover radii measurements, we thus adopt the median of the different turnover radii as our turnover radius and the standard deviation as its uncertainty. If one galaxy has only one turnover radii measurement, we adopt the typical scatter of 0.11 dex as its uncertainty. The measurements of the turnover features are listed in Table~\ref{tbl:turnover} for every turnover galaxy. 

Figure~\ref{fig:2d_maps} shows four examples of the turnover galaxies, analogously to Figure~\ref{fig:excluded}.  The first two are barred galaxies with regular morphologies, while the last two have slightly disturbed morphologies and are included in the pair catalogue. The first galaxy, MaNGA 8602-12701, has obvious breaks at $\approx$5$\arcsec$, and its central region is likely to be low-ionization nuclear emission-line region (LINER) in the BPT diagram.  The second galaxy, MaNGA 9487-12701, has similar profiles to the first one, but its central region is classified as star-forming according to the BPT diagram. The third galaxy, MaNGA 9507-12705 is an unbarred galaxy in a pair of Stage 3. It shows a break at 3\farcs5 in both \ewha\ and \dindex. The break radius is smaller than the bulge radius $\approx$5\arcsec. The last galaxy, MaNGA 8484-12703, is a barred galaxy in a pair of Stage of 2. It has a break in all three profiles, with a star-forming centre.

\subsection{Global properties of turnover galaxies}
\label{sec:global_properties}

\begin{figure*}
	\begin{center}
		\includegraphics[width=0.42\textwidth]{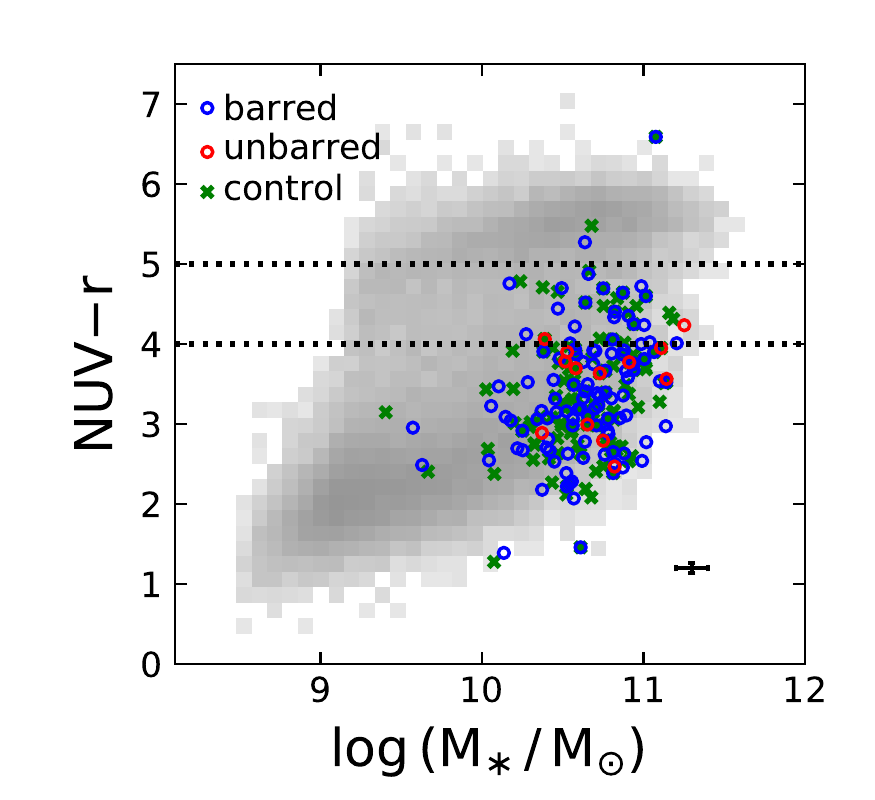}
		\includegraphics[width=0.42\textwidth]{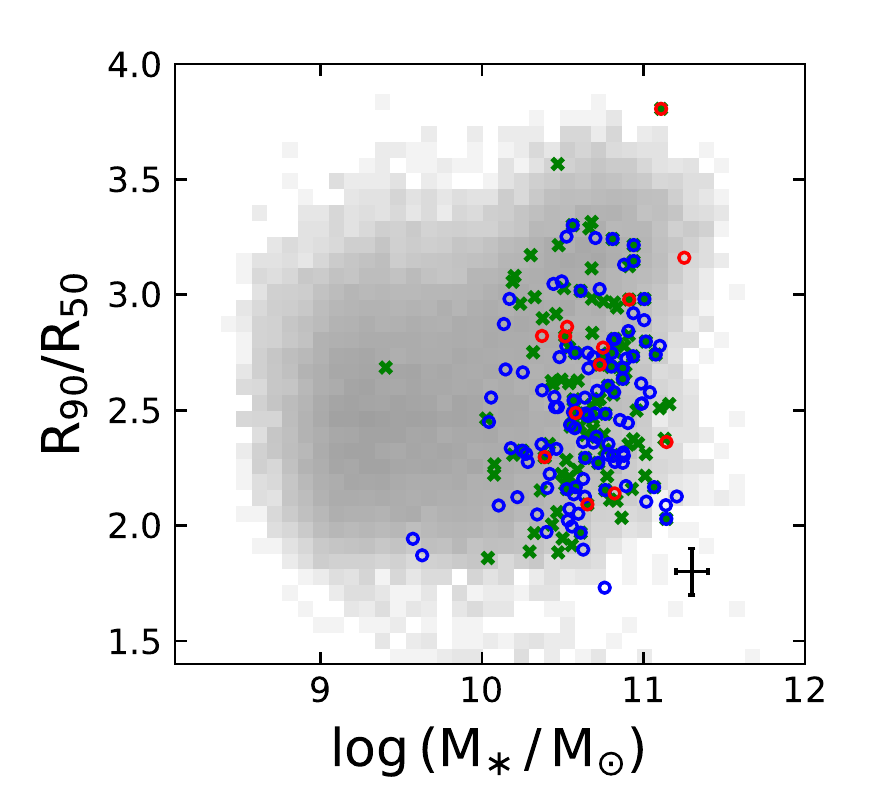}
		\includegraphics[width=0.42\textwidth]{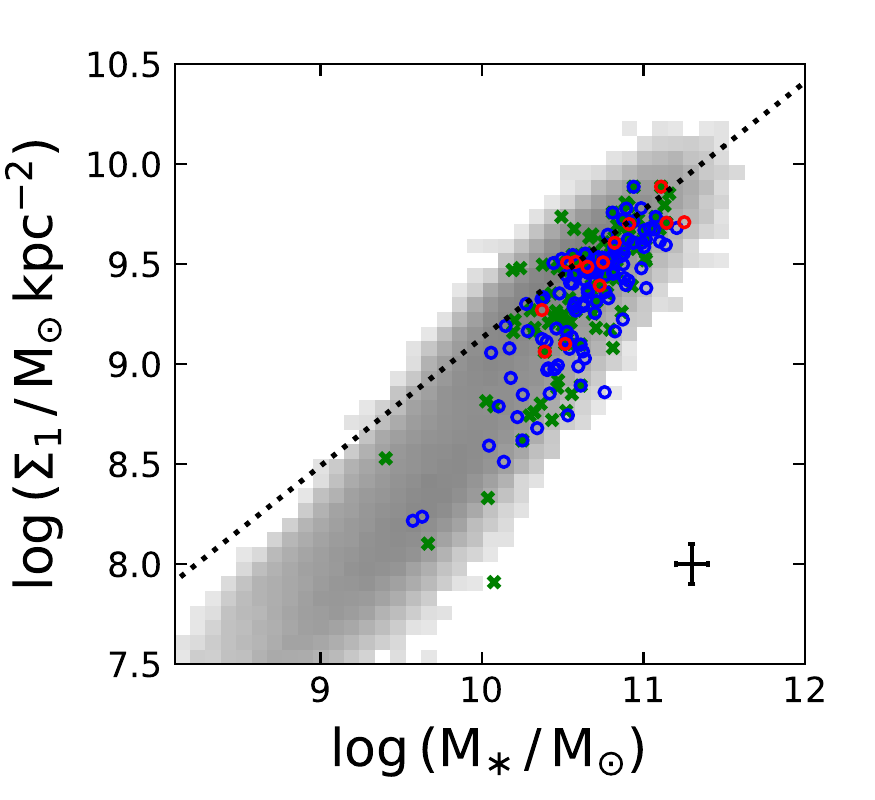}
		\includegraphics[width=0.42\textwidth]{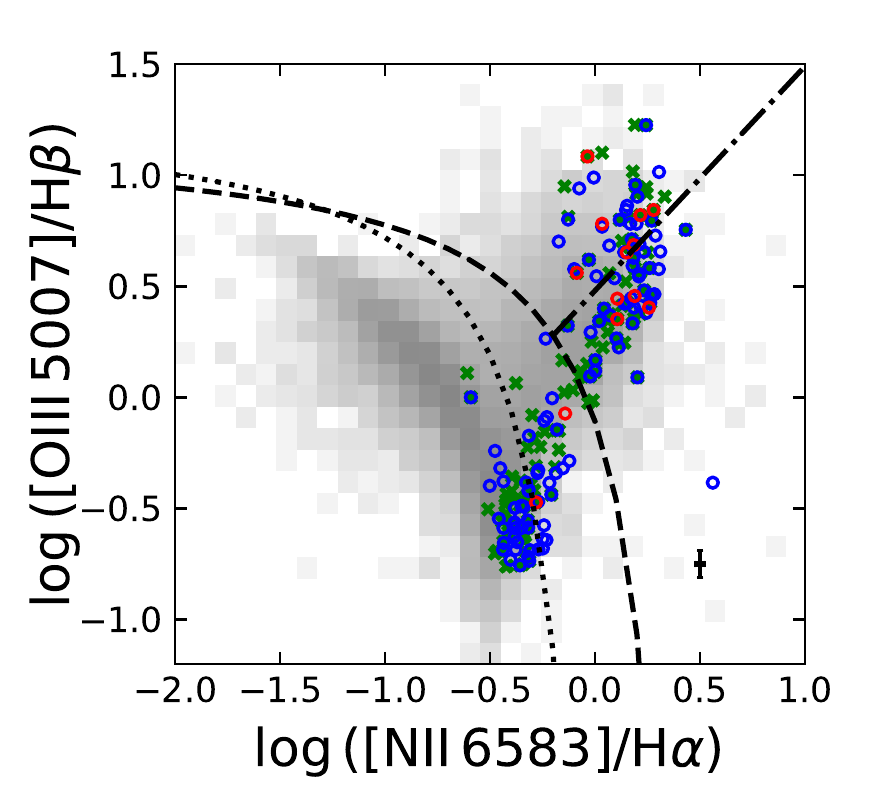}
	\end{center}
	\caption{Distribution of turnover galaxies in four diagrams: \nuvr--log$\,$(M$_{\ast}$/M$_{\odot}$), \rninety/\rhalf--log$\,$(M$_{\ast}$/M$_{\odot}$), log$\,$($\Sigma_{1}$/M$_{\odot}$kpc$^{-2}$)--log$\,$(M$_{\ast}$/M$_{\odot}$) and BPT diagram. Turnover galaxies are separated into barred (blue open circles) and unbarred (red open circles) galaxies. The control sample galaxies (green crosses) are matched in stellar mass and \nuvr\ colour. The volume-limited SDSS sample (greyscale) is shown as background. In the top-left panel, the black dashed lines are for NUV$-r=$ 4 and 5, commonly used to divide galaxies into blue-cloud, green-valley and red-sequence galaxies. In the bottom-left panel, the black dashed line indicates the $\Sigma_{1}$ threshold for quenching \citep{Fang-13}. In the bottom-right panel, the three black lines separate galaxies into the star-forming, composite, LINER and Seyfert categories \citep{Kauffmann-03, Kewley-06, CidFernandes-10}. Typical uncertainties are presented at the lower-right corner in each panel. }
	\label{fig:global_properties}
\end{figure*}

\begin{figure*}
	\begin{center}
		\includegraphics[width=0.33\textwidth]{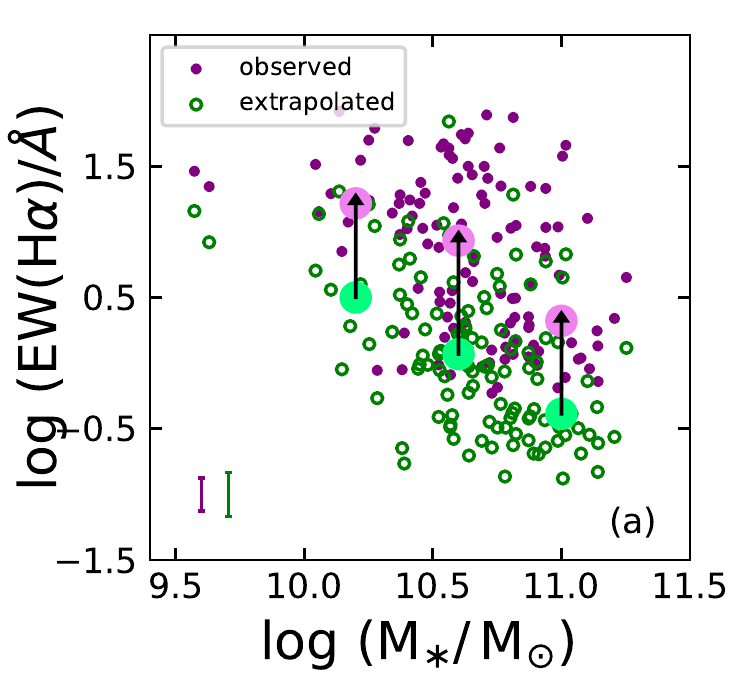}
		\includegraphics[width=0.33\textwidth]{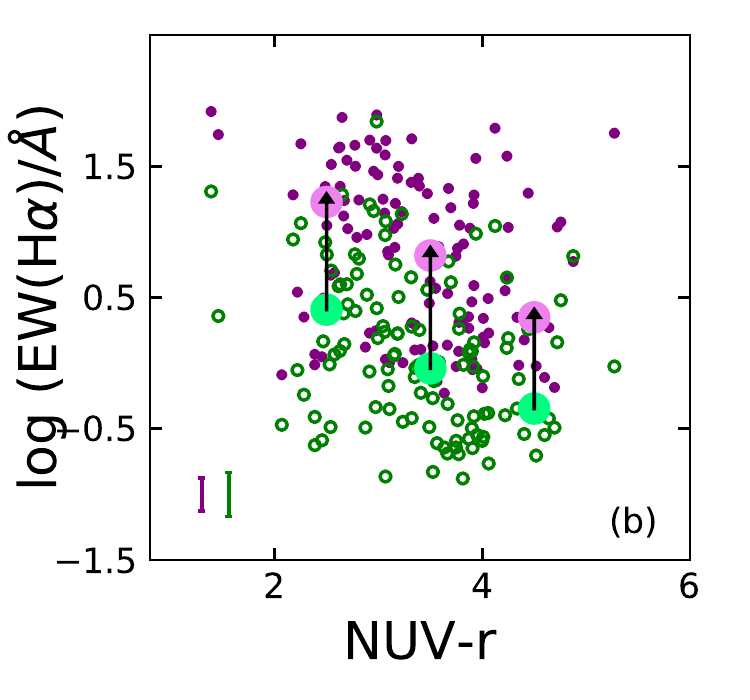}
		\includegraphics[width=0.33\textwidth]{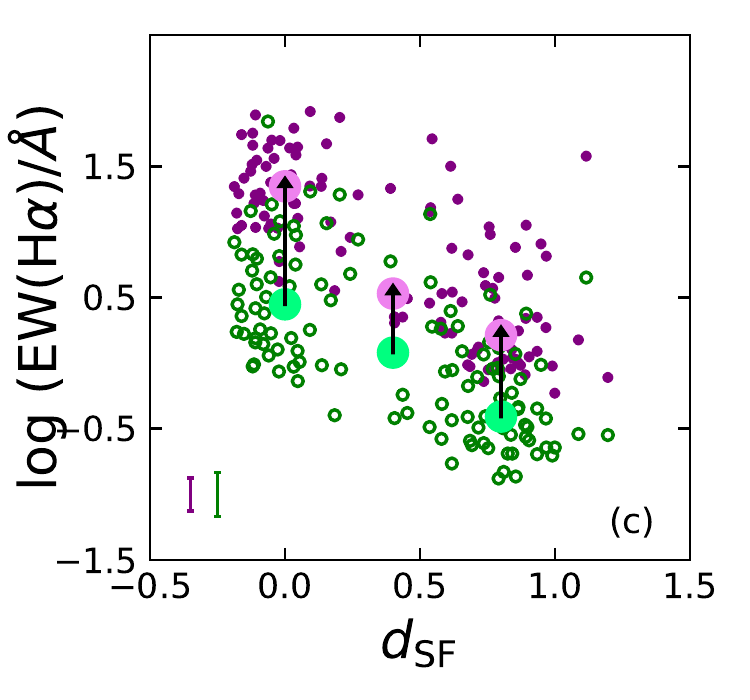}	
		\includegraphics[width=0.33\textwidth]{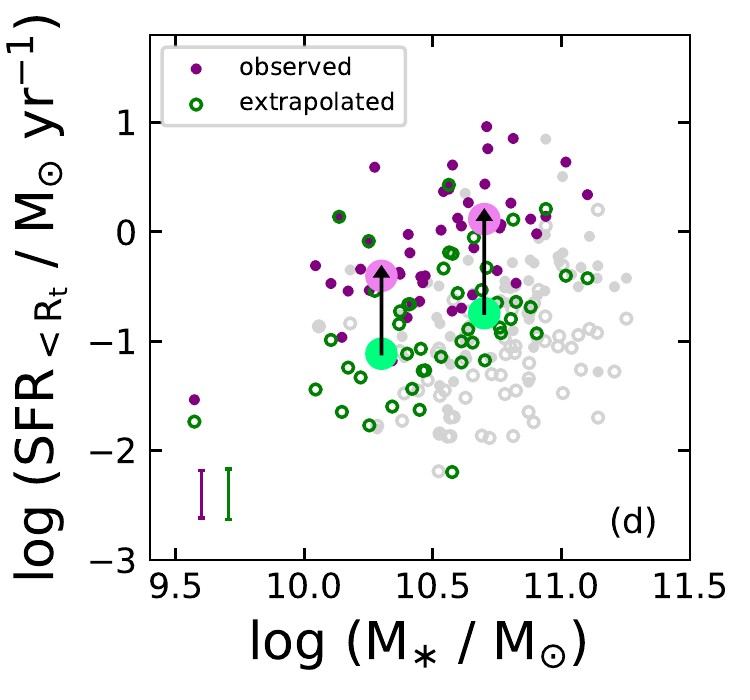}
		\includegraphics[width=0.33\textwidth]{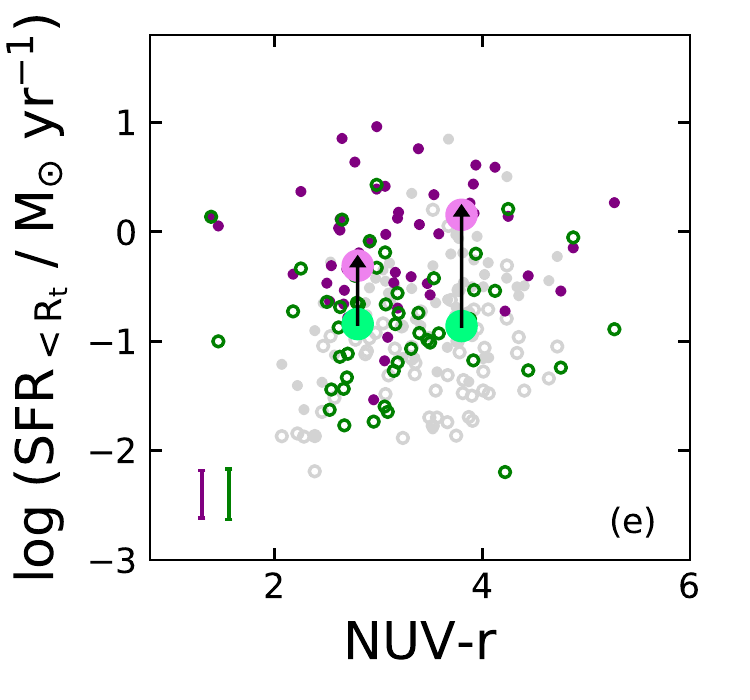}
		\includegraphics[width=0.33\textwidth]{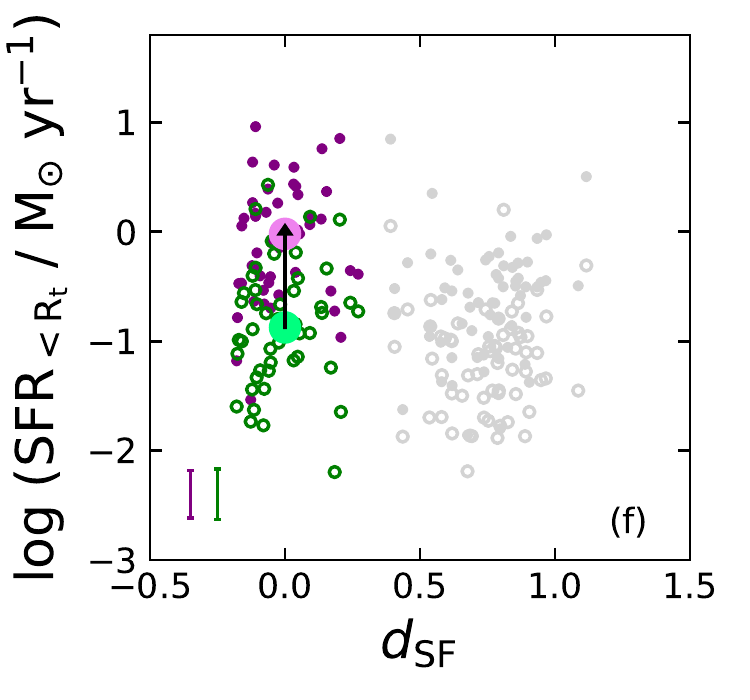}	
	\end{center}
	\caption{ $Top$: Central observed and extrapolated \ewha\ of turnover galaxies as a function of stellar mass (left), \nuvr\ colour (middle) and the distance from the SF sequence in the BPT diagram (right). The central observed and extrapolated \ewha\ are shown as purple solid circles and green open circles,  respectively. Error bars indicate their median errors. The similar but larger symbols indicate \ewha\ running averages at three intervals of stellar mass or \nuvr\ colour. The observed and extrapolated values at a given stellar mass or \nuvr\ colour are connected by a black arrow. $Bottom$: Observed and extrapolated SFRs within the turnover radii as a function of stellar mass (left), \nuvr\ colour (middle) and distance from the SF sequence in the BPT diagram (right). The SFR surface densities are calculated from the dust-corrected \ha\ fluxes, and then integrated from the centres to the turnover radii. Symbols are the same as in the top panels, except that galaxies in the LINER and Seyfert categories are shown in grey, as their \ha\ fluxes could be contaminated by nuclear activity.  }
	\label{fig:enhanced_lewha}
\end{figure*}


We start by examining the global properties of our turnover galaxies. For comparison, we also select a \textit{control} sample of galaxies from the MaNGA/MPL-7 by requiring the control sample galaxies and turnover sample galaxy to have similar stellar masses and  \nuvr\ colours. For this purpose, we make use of the stellar masses and \nuvr\ colours from the NSA catalogue. Typical uncertainties of stellar mass is about 0.1 dex\footnote{The uncertainty can be up to 0.3 dex due to systematic errors in different stellar population models and assumptions. In this work, the stellar masses are derived from $K$-correction fit, the relative rankings of stellar mass would not change much if they are estimated by different methods. }, \nuvr\ uncertainties are 0.06 dex \citep{Blanton-07}.  In practice, for each turnover galaxy, we search for a control galaxy from the MPL-7 with a stellar mass difference $\Delta$\lmstar/\msun\ $\leq$ 0.2 and a color difference $\Delta$(\nuvr) $\leq$ 0.3. We randomly select one of the many galaxies that meet these criteria, regardless of the presence or absence of a turnover feature. We note that the control sample includes 41 barred galaxies, i.e. it has a bar fraction of about 1/3, consistent with the typical strong bar fraction of galaxies in similar mass and colour ranges \citep{Barazza-Jogee-Marinova-08, Cheung-13}. In addition, the control sample includes 37 turnover galaxies, indicating a typical turnover galaxy fraction (30$\pm$4\%) for galaxies with $M_{\ast} > 10^{10}$ \msun\ and \nuvr\ $<5$. 

Table~\ref{tbl:numbers} lists the number of barred and unbarred galaxies, the number of galaxies at different pair stages, and the number of galaxies in different classes in the BPT diagram, for both the turnover galaxy sample and the control sample. The most striking difference between the turnover and control samples is that between their barred galaxy fraction, which is 89$\pm$3\% (108/121) and 34$\pm$4\% (41/121), respectively.  The high fraction of bars in turnover galaxies is expected, given the similarly high fraction of bars in the turnover galaxies from CALIFA (88$\pm$8\%, 15/17; see Paper I). This, again, strongly suggests a close relationship between the presence of a bar and a central star formation enhancement (as indicated by a turnover). 

In Figure~\ref{fig:global_properties}, we examine the global properties of our turnover galaxies by plotting them on four diagrams: \nuvr\ versus $\log M_{\ast}/M_{\odot}$, $R_{90}/R_{50}$ versus $\log M_{\ast}/M_{\odot}$, $\log \Sigma_{1}/M_{\odot} \rm kpc^{-2}$ versus $\log M_{\ast}/M_{\odot}$, and the BPT diagram.  The stellar surface mass  density within a radius of 1$\,$kpc, $\Sigma_{1}$, is calculated using the total light within the radius of 1$\,$kpc in $i$-band and stellar mass-to-light ratio $M/L_{i}$ from $g-i$ colour following \citet{Fang-13}. We adopt the typical uncertainties of color-based stellar $M/L$ ratio estimation are 0.1 dex \citep{Bell-03}, the typical uncertainties of $R_{50}$ and $R_{90}$ measurements are 4\% \citep{Blanton-11}. Emission line measurements and the corresponding uncertainties for the BPT diagram are taken from the MPA-JHU database\footnote{We refer to this as the MPA-JHU database, named after the Max Planck Institute for Astrophysics and the Johns Hopkins University where the measurements was developed. http://www.mpa-garching.mpg.de/SDSS/DR7/} \citep{Brinchmann-04}, based on the central 3$\arcsec$-fiber SDSS spectra. In Figure~\ref{fig:global_properties}, the barred and unbarred turnover galaxies are shown as blue and red circles, separately. The control galaxies are shown as green crosses. For comparison with the general galaxy population, we also select a volume-limited galaxy sample from the NSA, consisting of 35,070 galaxies with $r$-band absolute magnitude $M_r<-17.2$ and redshift 0.01 $<z<$ 0.03. The distribution of this sample is illustrated as the greyscale background in the figure.

In the colour-mass diagram (top-left panel of Figure~\ref{fig:global_properties}), galaxies can be divided into three regions: blue cloud, green valley and red sequence, as demonstrated by the volume-limited sample in the background. The turnover galaxies are found mainly in the blue cloud and green valley (\nuvr\ $\lesssim5$), with stellar masses $M_{\ast}$ ranging from $\sim$$10^{10}$ \msun\ to $\sim$$10^{11.5}$ \msun. After visually inspecting their SDSS images, we find most of the turnover galaxies to have morphological types from $Sab$ to $Sbc$, with very few later than $Sc$. This result suggests  that the turnover features indicative of enhanced star formation in galactic centres occur preferentially in massive spiral galaxies.  The ranges of stellar masses and colours found here for the turnover galaxies are consistent with those found from the CALIFA sample in Paper I.  By construction, the control sample covers similar ranges of stellar masses and \nuvr\ colours as the turnover galaxy sample.

The upper-right panel of Figure~\ref{fig:global_properties} shows the distribution of the different samples in the plane of concentration versus stellar mass, where concentration is quantified by the ratio of $R_{90}$ to $R_{50}$ in $r$-band ($R_{90}/R_{50}$).  After being matched with similar stellar mass, \nuvr\ colour and morphological type, the turnover sample is found to have similar concentrations with control sample. Majority of them have intermediate-to-low concentrations ($R_{90}/R_{50}$ $\lesssim$ 3). We further examine the bulge-to-total ratios (B/T) in turnover and control galaxies using the B/T measurements from the bulge-disc S\'ersic+exponential profiles decomposition catalog of \citet{Fischer-19}. Similar distributions are seen in turnover and control galaxies, with a median B/T of 0.2. We should point out that a three-component decomposition including bulge, bar and disc components would be preferable, given that the majority of turnover galaxies are barred. If bars are neglected in the decomposition, this usually results in an overestimation of the B/T ratio \citep{Gadotti-08}. Thus, one may expect a smaller B/T in turnover galaxies.

The central stellar mass surface density within the radius of central 1kpc, $\Sigma_{1}$, has been suggested to be a good indicator of the prominence of a central bulge \citep[e.g.][]{Fang-13, Luo-20}.  In the lower-left panel of Figure~\ref{fig:global_properties}, the black dashed line indicates the dividing line between quenched and unquenched galaxies proposed by \citet{Fang-13}:  at fixed mass, it has been found that most of the red (quenched) galaxies lie above this line. As can be seen in Figure~\ref{fig:global_properties}, a few (11$\pm$3\%, 14/121) turnover galaxies are above this line, consistent with their relatively blue \nuvr\ colours (\nuvr\ $\lesssim 5$) and relatively small concentration parameters ($R_{90}/R_{50}\lesssim3$). In contrast, more (17$\pm$3\%, 21/121) control galaxies are above the dividing line, suggesting the presence of a high-density central bulge. In addition, we have compared the central $r$-band surface brightnesses of the turnover and control sample galaxies,  finding the former to be brighter on average than the latter.  This implies a lower average mass-to-light ratio ($M/L$), thus younger stellar populations, in the centres of turnover galaxies compared to the centres of control galaxies. 

The lower-right panel of Figure~\ref{fig:global_properties} shows the sample galaxies in the BPT diagram.  The turnover and control sample galaxies have very similar distributions in this diagram, with galaxies distributed over all the different ionisation classes. However, when compared to the general galaxy population, the turnover galaxies (and the corresponding control galaxies) are limited to particular regions. In the star-forming region, the turnover and control sample galaxies are located in the region of low [OIII]/H$\beta$ and high [NII]/H$\alpha$. This can be understood from the relatively high stellar masses of the turnover galaxies. In the composite and AGN regions, the turnover and control galaxies are similarly found in the Seyfert, LINER and composite regions.  The number of galaxies of each type is listed in Table~\ref{tbl:numbers}.  We find that 25$\pm$4\%  (30/121) of turnover galaxies are classified as star-forming galaxies, 16$\pm$3\% (19/121) are composite,  32$\pm$4\% (39/121) are central LINERs and 21$\pm$4\% (25/121) are Seyferts. Turnover galaxies are likely to have slightly higher fractions for LINERs and Seyferts. We will return to these populations in Section~\ref{sec:discuss:tover_AGN_branch}. 

To summarise, central turnover features are found in massive spiral galaxies with $M_\ast $ $\gtrsim$ $ 10^{10}M_\odot$ and the majority of them are barred, falling in the blue cloud and green valley with \nuvr\ $\lesssim$ 5. Turnover galaxies have intermediate-to-low concentration parameters ($R_{90}/R_{50}$ $\lesssim$ 3) and relatively low central surface mass densities, suggesting relatively low bulge-to-total ratios. Their centres harbour all the different ionisation mechanisms in the BPT diagram, with slightly higher fractions of LINERs and Seyferts to those of the control sample.

\subsection{Enhanced star formation in turnover galaxies}
\label{sec:enhance_sf}

The central turnovers suggest young stellar populations are enhanced in the centres of turnover galaxies,  due to recent and/or ongoing star formation. To quantify these enhancements, we follow Paper I and obtain an \textit{extrapolated} value of  \ewha, \ewhda\ and \dindex\ for the central region of each galaxy. This is derived by extrapolating the linear fit to the radial profile from radii beyond the turnover radius down to the centre. The extrapolated parameters can be considered as the values expected if the central star formation were not enhanced. In the top panels of Figure~\ref{fig:enhanced_lewha}, we plot both the \textit{observed} (purple solid circles) and \textit{extrapolated} (green open circles) \ewha\ as a function of stellar mass (left panel) and \nuvr\ (middle panel). The similar but larger symbols indicate the \ewha\ running averages for three intervals of stellar mass or colour. The observed and extrapolated values at a given mass or colour are connected by a black arrow. As can be seen in the figure, the observed H$\alpha$ equivalent widths are enhanced by about 1 dex compared to the extrapolated ones, and this appears to be independent of stellar mass. The reddest colour bin is less enhanced, which may be reflecting the dependence of the central star formation enhancement on cold gas content at the centre.  However, given the relatively small number of galaxies in the reddest colour bin, this slightly weak enhancement should not be overemphasised.

Figure~\ref{fig:enhanced_lewha}c shows the observed and extrapolated \ewha\ as a function of  $d_{\rm SF}$, the effective distance from the star-forming sequence in the BPT diagram. Here, $d_{\rm SF}$ is calculated by using a set of curves parallel to but offset from the diagnostic line of \citet{Kauffmann-03} as a metric, to characterise the position of galaxies along the mixing sequence.  As can be seen, the centres of turnover galaxies can be broadly divided into two groups, i.e. those star-forming centres at low $d_{\rm SF}$ and those with AGN-like centres at high $d_{\rm SF}$ (with just a few galaxies in between, with $d_{\rm SF}\approx0.5$). The H$\alpha$ equivalent widths are enhanced by about one dex for star-forming galaxies, more than for galaxies with an AGN-like centre, that on average are enhanced by $\approx0.5$ dex. 

H$\alpha$ emission is a proxy for the star formation rate for star-forming regions. The data in the top panels of Figure~\ref{fig:enhanced_lewha} thus provide clear evidence for an enhancement of central star formation in turnover galaxies. To quantify the central SF enhancements more directly, we use dust-corrected \ha\ flux maps to infer SFR surface densities ($\Sigma_{\rm SFR}$), and then integrate the profile within the turnover region. Following common practice, the dust extinctions are calculated from the \ha/\hb\ ratios by assuming an intrinsic ratio of 2.86 and adopting a Milky-Way-like extinction law from \citet{Cardelli-89}.  The SFR calibration is given by \citet{Calzetti-13}:  SFR [\msun\,yr$^{-1}$] = 5.5$\times$10$^{-42}$ \Lha\ [erg\,s$^{-1}$]. The calibration assumes a \citet{Kroupa-01} stellar initial mass function (IMF) and a constant star formation timescale from 0.1 to 100 \msun. After deriving the SFR surface density profiles, we apply the same analysis as done previously for \ewha, \ewhda\ and \dindex, fitting the $\Sigma_{\rm SFR}$ profiles and deriving extrapolated $\Sigma_{\rm SFR}$ for the central regions. Then we integrate the profiles from the centres to the turnover radii to get the observed and extrapolated SFRs respectively, denoted as SFR$_{\rm <R_{t}}$.

The bottom panels of Figure~\ref{fig:enhanced_lewha} shows the observed and extrapolated SFRs of the central regions of the turnover galaxies as a function of $\log M_\ast/M_{\odot}$, \nuvr\ and $d_{\rm SF}$. The symbols/colours are the same as top panels, except that galaxies with LINER or Seyfert centres are shown in grey. We caution that the H${\alpha}$ emission in these galaxies could be contaminated by nuclear activity. The SFRs running averages indicated by the large symbols are calculated using SF galaxies only. Overall, turnover galaxies have SFR enhancements of about 0.5-1 dex, similar to the trends seen in \ewha.

\begin{figure*}
	\begin{center}
		\includegraphics[width=\textwidth]{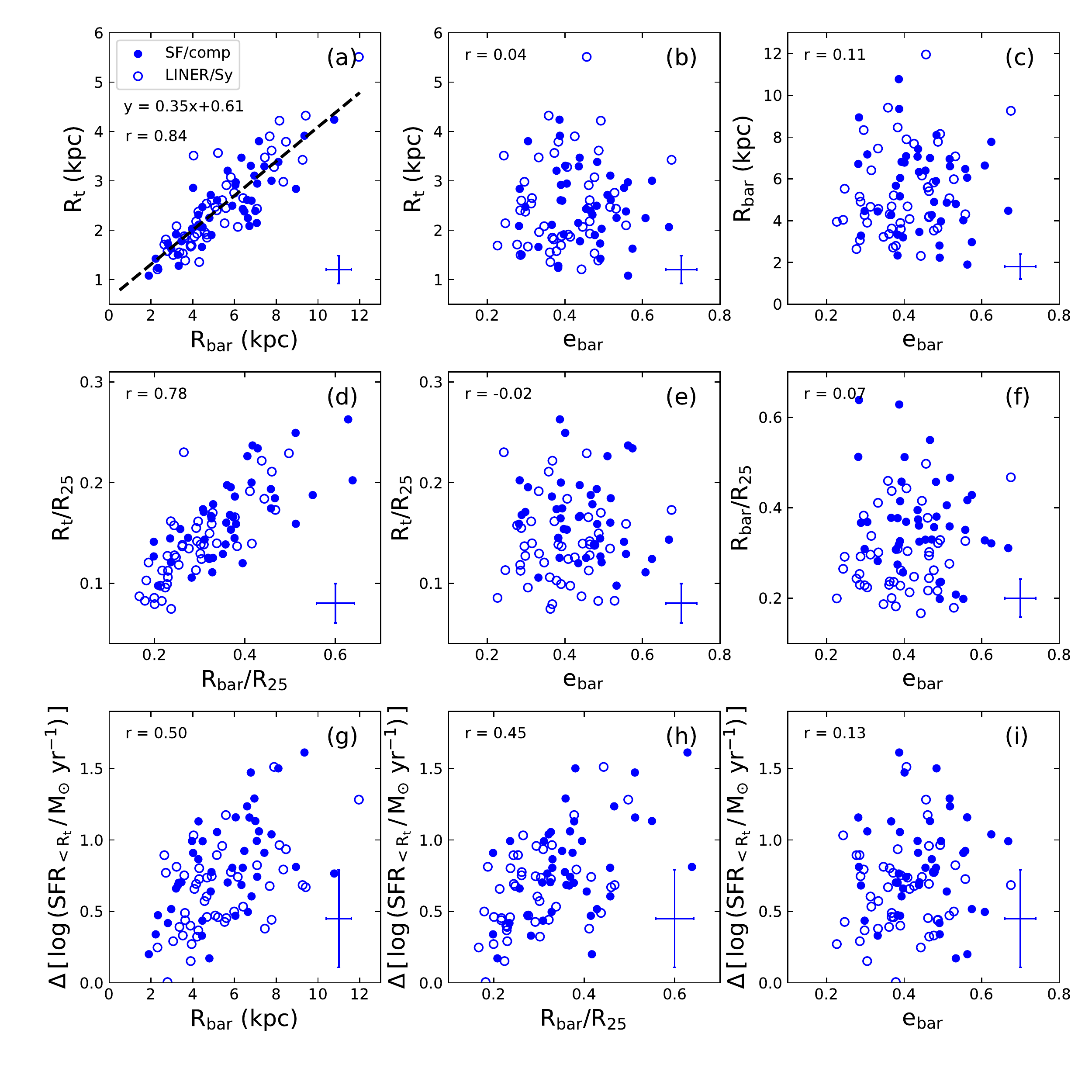}
	\end{center}
	\caption{Turnover quantities as a function of bar properties. Bar properties are described by the bar length (\rbar), normalised bar length (\rbar/$R_{\rm 25}$) and bar ellipticity (\ebar) along the abscissa. Turnover features are characterised by the turnover radius \rt, the normalised turnover radius  \rt/$R_{\rm 25}$ and integral SFR enhancement within \rt\ ($\Delta$\,[log\,SFR$_{\rm <R_{t}}$/M$_{\odot}$yr$^{-1}$] ) along the ordinate. Only barred turnover galaxies are shown. Data points are separated into SF (solid blue circles) and AGN (open blue circles) according to their locations in the BPT diagram. We caution that the SFR estimate in LINER/Syfert galaxies could be contaminated by nuclear activity. The dashed line in the top-left panel shows a linear fit between turnover radius and bar length. The Spearman correlation coefficients $r$ are listed at the top-left corner. Typical uncertainties are presented at the lower-right corner. }
	\label{fig:tover-bar}
\end{figure*}

\subsection{Relationship between turnovers and bars}
\label{sec:correlation_with_bar}

Now we examine the relationship between the turnover features (including the turnover radius \rt\ and the SFR enhancement within the turnover radius $\Delta$\,[log\,(SFR$_{\rm <R_{t}}$/\msun yr$^{-1}$)]) and bar properties (including the bar length \rbar\ and bar ellipticity \ebar). The results are shown in Figure~\ref{fig:tover-bar}. In each panel the turnover barred galaxies of star-forming and composite types and those of LINER and Seyfert types are shown as solid blue and empty blue circles, respectively. In some panels, we normalise the bar length and turnover radius by $R_{25}$, the radius at which the $r$-band surface brightness is 25 mag arcsec$^{-2}$. 

The top-left panel of Figure~\ref{fig:tover-bar} presents the tightest correlation --- an increasing of the turnover radius with increasing bar length, a trend holding for both SF/composite galaxies and AGN. This correlation can be well described by a linear function, \rt\ $=$ (0.35$\pm$0.02)\rbar+(0.61$\pm$0.13), plotted as a dashed line in the panel. A similar correlation holds, albeit with more scatter, when \rt\ and \rbar\ are scaled by $R_{25}$  (Figure~\ref{fig:tover-bar}d).  The star formation enhancement within the turnover region, as quantified by $\Delta$\,[log\,(SFR$_{\rm <R_{t}}$/\msun yr$^{-1}$)] is also positively correlated with \rbar\ and \rbar/$R_{25}$  (Figure~\ref{fig:tover-bar}g,h), but with with larger scatter than the correlations between turnover radius and bar length.

Unlike bar length, bar ellipticity shows no correlation with turnover radius or SFR enhancement.  Both bar length and bar ellipticity are commonly used as proxies of bar strength, because of their correlations with gravitational bar torques \citep[e.g.][]{Combes-Sanders-81, Block-01}. Bar length and ellipticity are indeed correlated, as shown by \citet{Wang-12} with a large sample of barred galaxies from SDSS. However, the turnover galaxies in our sample show no correlation between \rbar\ and \ebar, as can be seen in Figure~\ref{fig:tover-bar}c. Therefore, the lack of correlation between the turnover features and bar ellipticity is likely due to the limited size of our sample, or it may be suggesting that bar length is really more fundamental to the turnover feature. To clarity this, one would need a larger sample of turnover galaxies as well as theoretical studies (such as numerically simulating the co-evolution of galactic centres and bars). Besides, the measurement of bar ellipticity using the ellipse fitting technique would be affected by B/T, bar strength and other features like spiral arms or rings. We matched our sample with the catalogue in \cite{Kruk-18} in which they considered bar in photometric decomposition. 44 galaxies are included in their measurements. We still found no obvious correlations between turnover radius versus ellipticity or B/T in their catalogue. Hopefully this can be examined again with larger samples in the future.


\section{Discussion}
\label{sec:discuss}

\subsection{Central versus global star formation enhancements}

Central turnovers were identified by comparing the observed and inward extrapolated values of the three parameters closely related to recent star formation that we considered.  These turnovers reflect {\it relative} SF enhancements with respect to the pre-existing underlying old populations at the galaxy centres. Therefore, the central turnover galaxies selected in this way are expected to have recently experienced  bulge growth and/or rejuvenation (Thomas \& Davies 2006; Coelho \& Gadotti 2011; Mendez-Abreu et al. 2014; Chown et al. 2019).


In the literature, there have been various attempts to estimate SFR enhancements in the central regions of galaxies, driven by instabilities induced by bars or tidal interactions. For instance, \cite{Li-08} quantified the enhancements of SFR as a function of projected distances to neighbouring galaxies, finding the SFRs can be enhanced by a factor of 2$-$3 at scales comparable to the sizes of individual galaxies. \citet{Ellison-11} estimated the average central $\Delta$\,log\,SFR of barred galaxies relative to unbarred galaxies at fixed stellar mass, finding the former to be 0.2$-$0.3\,dex higher. \citet{Wang-12} calculated the central-to-total SFR ratios of a sample of barred galaxies and a carefully-selected sample of control galaxies, finding the average SFR ratio of galaxies with strong bars is about 0.2$-$0.3 dex higher than that of the control galaxies (see Table 1 of their paper).

The typical SF enhancement of our turnover galaxies is $\approx$1.0 dex (see Figure~\ref{fig:enhanced_lewha}), higher than those previously found (0.2$-$0.4 dex as described above).  We argue that this difference can be mainly attributed to the different definitions of star formation enhancement. In our case, the enhancement is quantified for the central region of the each galaxy, comparing the observed and expected (extrapolated) SFRs. In fact, in a recent paper, \citet{Chown-19} estimated the enhancements of the central SFRs of barred galaxies in the same way as in our work, also finding a high enhancement of $\approx$1 dex (see their Fig. 5).  In previous studies, the enhancement of each galaxy was calculated by comparing the central SFR with either the total SFR of the whole galaxy \citep[e.g.][]{Wang-12} or the central SFR of control galaxies \citep[e.g.][]{Li-08, Ellison-11}. In fact, we re-calculated $\Delta$\,log\,SFR by comparing the central SFR of each turnover galaxy with the central SFR of its control galaxy, and found an average SFR enhancement of 0.28 dex, thus consistent with previous studies.

Nevertheless, we would like to emphasise that the turnover galaxies studied here are not necessarily the same objects as those studied in previous studies, that usually present globally higher SFRs, with peak SFRs in the centre and weaker SFRs at larger radii.  By investigating the spatially-resolved SFR profiles, galaxies above (or below) the star formation main sequence (SFMS) are found to have enhanced (or suppressed) star formation at all galactocentric radii \citep[e.g.][]{Eillson-18, Guo-19}, though with stronger effects at the centre. This can be interpreted as variations of the overall gas inflow rates \citep[e.g.][]{Lilly-13, Wang-19}. However, our turnover galaxies are expected to have star formation enhanced at their centres comopared to the outskirts of the bulge or bar regions. According to our sample selection criteria, turnover galaxies do not necessarily have globally higher star formation rates compared to the SFMS. Therefore, the turnover features are related to radial gas inflows to the galactic centres, rather than enhancements of the overall gas inflow rates. 

Besides, instead of considering central star formation enhancement, one could think about turnover behavior in another way: the central region is surrounded by a ``desert'' ring, where the star formation has been suppressed \citep{James-Percival-18}. According to the bar evolution scenario proposed in \citet{James-Percival-18}, the star formation enhancement would be an initial response after the radial gas inflow, after that, the bar would suppress SF and form a desert ring. In this study, we define turnover galaxies according to their radial profiles. Thus our sample could include SF enhancement at the beginning phase, as well as objects at the later phase that the surrounding area is being suppressed, while the central SF still exist. The presence of a star formation desert potentially increases the contrast between central and surrounding region in the radial profiles, which could bias the ability to detect turnover features towards strong barred galaxies. 

\subsection{Lack of ``turnover'' galaxies below $M_\ast\sim10^{10}$ M$_\odot$}
\label{sec:discuss:lack_of_low_mass}

It is interesting to note that the turnover features identified here are mostly found in relatively high-mass galaxies, with stellar masses above 10$^{10}$ \msun.  Although adopting a different definition of star formation enhancement, the earlier study of barred galaxies in SDSS by \citet{Ellison-11} also found a lack of star formation enhancement at stellar masses below 10$^{10}$ \msun. The lack of central turnover features in low-mass galaxies indicates a lack of central star formation enhancements, that probably is not unexpected for low-mass galaxies. First, most low-mass spirals do not have a clear bulge structure or are  even bulgeless. Furthermore, in most cases, the low-mass galaxies in our sample are actively forming stars (and are thus entirely dominated by young stellar populations) according to their \ewha, \ewhda\ and/or \dindex\ profiles, that are typically flat or slightly negative, and could thus be well described by a single slope. Therefore, low-mass galaxies may be globally enhanced, i.e. with star formation enhancement at all radii \citep[e.g.][]{FraserMcKelvie-19}. In this work, we only considered the central regions of our galaxies.  In a following paper we will examine the characteristic radial profile of star formation enhancement (or suppression) as a function of galaxy mass and other properties. 

\subsection{Role of bars}
\label{sec:discuss:role_of_bars}

We find 89$\pm$3\% of the turnover galaxies are barred, a very high fraction compared to the bar fraction of 34$\pm$4\% in the control sample. A similarly high fraction of 88$\pm$8\% was found in our previous work \citep{Lin-17} from the CALIFA survey. Combined with the finding of \citet{Chown-19} on centrally-concentrated molecular gas in turnover galaxies, our results thus reinforce the conclusion that bars must play a dominated role in enhancing central star formation in (barred) galaxies, by driving cold gas inflows.

In this work we have further examined the correlation of turnover features with bar properties, as quantified by bar length and ellipticity.  It is striking that turnover radius is tightly correlated with bar length, the turnover occurring at about 1/3 of the bar length. This is, again, strongly suggesting that bars are indeed the main driving factor of turnover features.  In a few previous studies, similar turnover positions have been found, e.g. \citet{Seidel-15} and \citet{Seidel-16} identified breaks in the radial profiles of kinematic features and stellar population indices of barred galaxies. Those authors found the breaks usually happen at 0.13$-$0.2 \rbar. The turnover radii in our galaxies are slightly larger, likely due to the fact that the turnover features in our work are traced by young stellar populations. As the central bulges grow, one would expect to find younger populations at larger radii, consistent with the inside-out formation of bars as predicted by numerical simulations \citep[e.g.][]{Athanassoula-16}. A recent study that looks at the stellar populations in the barred galaxies also find a similar break at $\approx$0.3 \rbar\ \citep{Neumann-20}.

We can also compare our bar classifications with the barred galaxies of the Galaxy Zoo (GZ) catalogue \citep{Willett-13}. In the MPL-7 sample,  $\approx$ 90\% galaxies can be matched with the current GZ catalogue.  Using the criteria bar probability \textit{p\_bar} > 0.5 and non-edge-on probability \textit{p\_edge} < 0.5, 269 objects are  selected as barred galaxies. The bar fraction (19$\pm$1\%) is consistent with our own (18$\pm$1\%). Comparing the two barred samples, 192 objects ($\approx$70\%) overlap. The differences are mainly due to weak bars, whose identification is somewhat ambiguous and subjective, both GZ and our classification miss or mis-identify some weak bars, and we identified more lenticular objects as barred galaxies. Nevertheless, regarding our turnover sample, if we use the GZ bar classifications, there are 89 barred turnover galaxies, 26 unbarred turnover galaxies and 6 galaxies lacking a classification. These numbers yield a similarly high bar fraction of 77$\pm$4\% among turnover galaxies. So our results would not change much if we used GZ bar classifications.

\subsection{Role of interactions/mergers and local density}
\label{sec:discuss:correlation_with_environment}

\begin{figure}
	\begin{center}
		\includegraphics[width=0.4\textwidth]{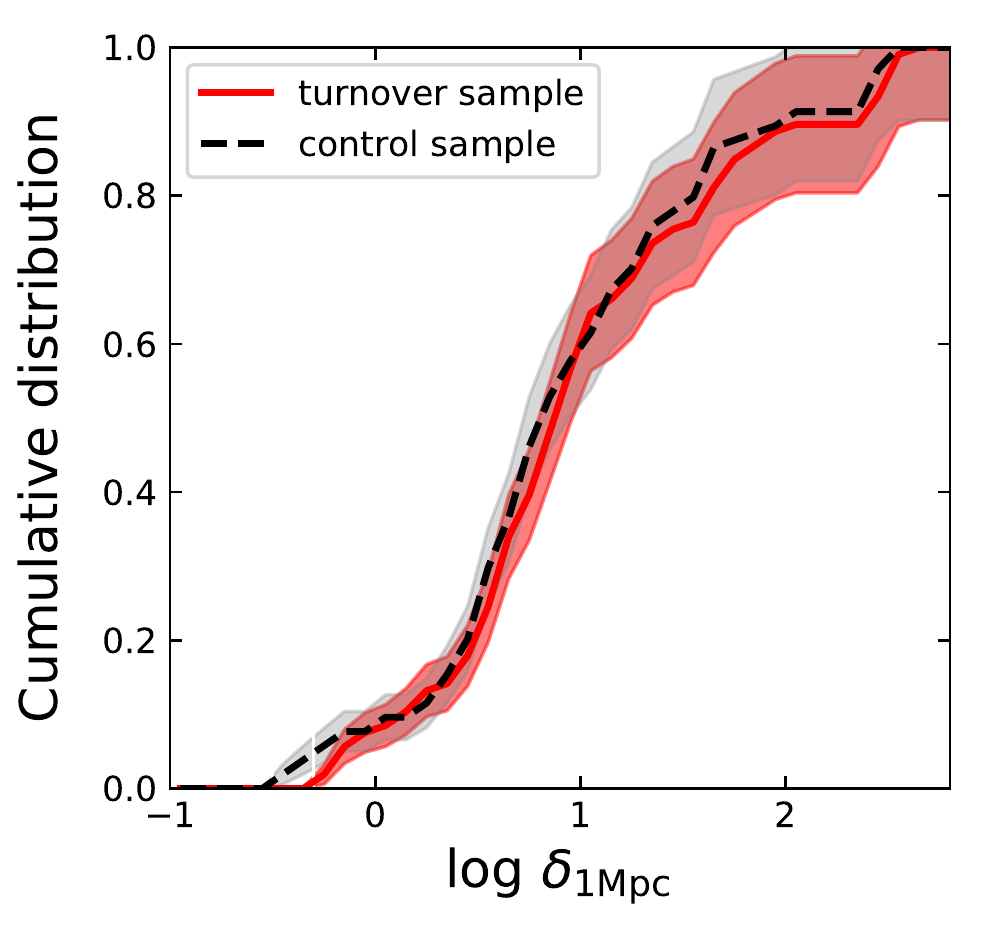}
	\end{center}
	\caption{Cumulative fraction of local density in 1 Mpc for turnover galaxies (black dashed line) and control sample galaxies (red solid line). The shaded areas represent the uncertainties computed by assuming the Poisson distribution. The K-S test probability is 0.15, suggesting no significant difference between the two samples. }
	\label{fig:local_density}
\end{figure}

In addition to bar-driven instabilities, tidal interactions with close companions are also known to be able to drive cold gas from the outer parts of discs to galactic centres, thus leading to similar star formation enhancements in the central regions of galaxies \citep[e.g.][]{Li-08, Ellison-08}.  For instance, a recent MaNGA-based study by \citet{Pan-19} revealed significant star formation enhancements in paired galaxies with a separation less than 20 kpc, an effect that is strongest at the galaxy centres. We have therefore also examined the correlation of our turnover galaxies with galaxy pairs/mergers, but find no obvious link. Of the 121 turnover galaxies, only 20 are in a pair/merger according to the classification of \citet{Pan-19}. This fraction (17$\pm$3\%) is consistent with the fraction of pair/merger galaxies in the control sample (24/121).  On the other hand, of the whole 228 pairs/mergers galaxies in our parent sample, there are only 20 galaxies are identified as turnover galaxies, the fraction (9$\pm$2\%) is even lower.

In our turnover galaxies, except the 108 barred galaxies, there are 16 turnover galaxies with no bar. One may expect the 16 unbarred turnover galaxies to be dominated by galaxy pairs/mergers, so that bars and tidal interactions combined can completely explain the central star formation enhancements of our sample galaxies. Indeed, in a recent study, \citet{Chown-19} examined the central star formation enhancement in a sample of 58 nearby galaxies from CALIFA, finding the enhanced star formation are either barred (13/19, 68$\pm$11\%) or in pairs/mergers (6/19, 32$\pm$11\%). However, this is not the case in our sample, where we find a similarly high bar fraction in the turnover galaxies (89$\pm$3\%), but only 4 of the 13 unbarred turnover galaxies are associated with pairs/mergers (31$\pm$13\% in our sample versus 100\% in theirs). Most (16/21) of our turnover galaxies with galaxy pairs/mergers are also barred.

In conclusion, in contrast to our initial expectation, galaxy interactions/mergers are not obviously related to the central star formation enhancement as indicated by the central turnover features. This is true for the majority of the turnover galaxies, regardless of whether they are barred or unbarred. This result may be caused by multiple reasons.  First, several paired galaxies are dry mergers, thus lacking cold gas to support the central star formation. We found that about 13\% (30/228) of the paired galaxies have \nuvr\ $>$ 5, with little cold gas to drive and/or support central star formation. Second, as pointed out in Section~\ref{sec:discuss:lack_of_low_mass}, low-mass galaxies usually have weak or no bulge component, showing no turnover features in their centers, although their star formation may be globally enhanced. The rest paired galaxies (198/228) with \nuvr\ $<$ 5, more than half (54$\pm$3\%, 124/228) have stellar masses less than 10$^{10}$ \msun. For these galaxies, interactions/mergers probably have a more important role in enhancing star formation over large areas of the galaxies, rather than only driving central star formation enhancements. We indeed see several paired galaxies in which enhanced star formation extends to large radii, far beyond the central regions. In our work, they are not selected as turnover galaxies as the turnover radii are defined to be  smaller than the size of the bulge or bar. Third, recent studies suggest that star formation enhancements are significant only when the paired galaxies are close enough. For instance, \citet{Pan-19} found significant star formation enhancement when pair separation is less than 20 kpc. Finally, other parameters like the mass ratios of the paired galaxies, their merging geometries and the gas fractions should also affect the resulting star formation enhancements. 

We have also examined the environments of our galaxies over larger physical scales. First, using the group catalogue of \citet{Yang-07}, we find that the ratio of central and satellite galaxies in the turnover galaxy sample is very similar to that of control sample, with a central galaxy fraction of $\approx$70\% in both cases. Next, based on the density field of the local Universe reconstructed by \citet{Wang-16},  Figure~\ref{fig:local_density} shows the cumulative distribution of the local mass density ($\delta=\rho/\bar{\rho}$) measured over a scale of 1\,Mpc for both the turnover galaxies and the control sample galaxies.  The two samples show very similar cumulative density distributions, with a Kolmogorov-Smirnov (K-S) test probability larger than 10\% indicating no significant difference between the two distributions. Therefore, we conclude that the large-scale environment is not obviously linked to the turnover features of our galaxies.

\subsection{Role of cold gas}

Central star formation enhancements are expected to be associated with higher cold gas contents in galactic centres. In fact, \citet{Chown-19} jointly analysed the integral-field spectroscopy from CALIFA and CO line imaging from the Extragalatic Database for Galaxy Evolution (EDGE) survey for a sample of nearby galaxies. They found the central enhancements of star formation rate to be positively correlated with the concentrations of molecular gas, an effect only present in barred galaxies and those in pairs/mergers. This provides direct evidence that cold gas inflows due to bar- or merger-driven instabilities enhance star formation in the central regions of galaxies. On the other hand, \citet{Ellison-20} studied the CO content of MaNGA galaxies with a central starburst, finding that the  elevated star formation rates in their centres are instead driven by enhancements of the star formation efficiency, defined as the star formation rate divided by the \htwo\ mass. Nevertheless, they also found lower gas-phase metallicities in the central regions of starburst galaxies compared to their outer parts, consistent with a metal-poor gas inflow scenario.

In our previous work \citep{Lin-17}, a weak correlation was also found between the turnover features and the \hi\ gas mass fraction of their host galaxies. Here we attempted to examine the \hi\ gas mass of our MaNGA galaxies using both the ALFALFA survey \citep{Haynes-18} and the \hi\ follow-up observations being carried out by the MaNGA team \citep{Masters-19}, which is aiming to observe all MaNGA galaxies with $z<0.05$. We thereby obtained \hi\ gas masses for 90 turnover galaxies, but there is no significant correlation of neither turnover radius or $\Delta$\,[log\,(SFR$_{\rm <R_{t}}$/\msun yr$^{-1}$)] with the \hi\ masses. We also compared the \hi\ fraction in turnover barred and non-turnover barred galaxies, and we find no obvious difference either.  It is thus very likely that the weak correlation found in our previous work is not real, but was due to the small sample size. Having said that, the sample from MaNGA is larger but still rather limited. Further analysis of the \hi\ gas content of our galaxies is beyond the scope of this paper, and we leave it to future works when larger samples of \hi\ observations are available.

Nevertheless, even with a more complete sample, the absence of a correlation between central star formation enhancement and \hi\ gas mass may not be surprising, for the following two considerations.  First, it is widely believed that atomic gas is less closely related to star formation than molecular gas \citep{Bigiel-08, Schruba-11}.  Second, the cold gas is expected to be driven towards the centres of the galaxies, but the global gas contents are not necessarily increased.  This is indeed what was found by \citet{Chown-19} considering the spatially-resolved CO maps. Unfortunately, spatially-resolved \hi\ maps are not yet available for our galaxies. In a sample with spatially-resolved \hi, selected to have both strong bars and significant \hi\ content, \citet{Newnham-20} demonstrated that central \hi\ holes are common, but not ubiquitous in barred galaxies. In that sample, the presence or absence of a \hi\ hole correlated with total \hi\ gas fraction, as well as the global star formation properties, with barred galaxies without central \hi\ holes more likely to have the highest gas fractions and ongoing global SF.

\subsection{Turnover galaxies with central LINER/Seyfert emission}
\label{sec:discuss:tover_AGN_branch}

It is notable that nearly half of our turnover galaxies are located in the AGN region of the BPT diagram, including 32$\pm$4\% (39/121) LINERs and 21$\pm$4\% (25/121) Seyferts. This appears to be in conflict with our expectation that the central regions of these galaxies are dominated by recent/ongoing star formation, as indicated by the turnover of \ewha, \ewhda\ and \dindex.

The origin of LINER emission has been widely discussed in recent years. Rather than connecting it with central low ionisation nuclear activity as traditionally-assumed, recent studies suggest LINER-like emission may be produced by evolved stars, in particular post-AGB stars \citep{Stasinska-08, CidFernandes-11, Belfiore-16}. This again appears to be in conflict with our expectations. Is it possible that the H$\alpha$ emission produced by star formation is mixed with diffuse LINER emission?  To answer this question, we have examined the locations of our turnover regions on the spatially-resolved star formation main sequence \citep{Rosales-Ortega-12, Sanchez-13}. As suggested by \citet{Hsieh-17}, the location on the spatially-resolved main sequence diagram may indicate whether the H$\alpha$ emission is powered by young or old stellar populations. We find that most of the turnover regions lie between the star forming sequence and the quiescent sequence.  This supports our conjecture that the turnover regions  are a mixture of underlying old stellar populations and recently-formed young populations. 

In addition, \citet{Belfiore-16} separated LINER-like galaxies into central-LIER (cLIER) and extended-LIER (eLIER) according to the location of the LINER emission within each host galaxy. Our LINER-like turnover galaxies are likely to be cLIERs based on their morphologies. The central \dindex\ of these cLIERs range from 1.6 to 1.9, systematically lower than the \dindex\ of eLIERs (see Fig. 12 in \citet{Belfiore-16}), consistent with our suggestion that their central regions contain significant fractions of young stellar populations.

For the turnover galaxies with Seyfert-type emission, the \ha\ enhancement could actually be contributed by AGN activity.  Communicating privately with colleagues working on AGN sources, we note that there are no any broad emission line in these galaxies. If this is true, these Seyfert-type turnover galaxies should be low-luminosity AGN. We find that most of them have turnovers in both \ewha\ and \dindex, while only 5 of them have only one \ewha\ turnover. Since AGN can contribute to \ha\ emission but not to \dindex, the associated \dindex\ turnovers indicate that these could be systems with AGN surrounded by circumnuclear star formation. The Seyfert-type fraction in the turnover sample (21$\pm$4\%, 25/121) is slightly higher than that in the control sample (13$\pm$3\%, 16/121), implying that the physical processes (e.g. bar-driven instability) driving central star formation also tend to boost nuclear activity. Given that the turnover galaxies are predominately barred, the association of the turnover feature with Seyfert-type emission also suggests a physical link between the presence of bars and nuclear activity.

Although many studies agree that bars indeed cause an inflow of gas toward the central regions of galaxies \citep[e.g.][]{Regan-Teuben-04, Sheth-05, Hunt-08} as summarised in the Introduction, the connection between bars and nuclear activity is still under debate. A few high-resolution observations see fueling of nuclei through bars \citep[e.g.][]{Fathi-13}, while many statistical studies do not find a different incidence of AGN activity associated with a bar \citep[e.g.][]{Ho-Filippenko-Sargent-97, Hao-09}. Numerical simulations suggest that additional effects, such as a secondary bar or a nuclear ring, are required to drive the gas further inwards \citep[e.g.][]{Regan-Teuben-04}. Recent high resolution observations also suggest that there are many different ways to fuel galactic nuclei \citep[e.g.][]{Hunt-08, Gadotti-19}. To figure out the connection between central turnover features and nuclear activity, high-resolution observations of gas and kinematics will be needed.

\subsection{Future perspectives}

In this work, we focus on the turnover feature happened within the inner region where a bulge is usually present. However, the star formation enhancement can be more extended and beyond the entire bulge. For example, the first galaxy (7962-12704) in Figure~\ref{fig:excluded} indeed shows a break in \ewha, \ewhda\ and \dindex\ profiles. 
The break happens at $\sim$4$\arcsec$, while the central bright core (likely a pseudo-bulge) is much smaller, $\sim$1$\arcsec$, according to the SDSS image. Therefore, the ``turnover'' seen in the profiles cannot be a feature of the central core. Rather, the break happens at the transition between the inner region and the spirals. The whole inner region is young and without break in the profile. Besides, we also see some interacting galaxies which have extreme star formation across the entire galaxy. There is no turnover in the profile either because the entire galaxy is young and the profile can be fitted by single slope. 

These phenomena are also interesting by themselves. They probably result from similar physics as turnover galaxies, such as non-axisymmetric-induced gas inflow through spiral arms or tidal interactions with close companions. Different selection criteria are needed to search for these galaxies, which can be an additional topic for future studies. 

Many nuclear structures (such as nuclear ring, secondary bar and nuclear spiral) are likely to be built by the infalling gas and intense star formation. They help build up the central bulge and also play an important role in the secular evolution. A number of studies had been done from both theoretical and observational \citep{Shlosman-89, Piner-95, Knapen-95, Comeron-10, Cole-14, Seo-19} to understand their physical properties. High spatial resolution observations from MUSE or ALMA are essential for fully understanding the formation and evolution of these substructures.

\section{Conclusions}
\label{sec:conclusion}

In this work we analysed spatially-resolved maps and radial profiles of \ewha, \ewhda and \dindex\ for a sample of $\approx1400$ nearly face-on galaxies at $z<0.05$, using integral field spectroscopy from the current sample (MPL-7) of the MaNGA survey.  We identified 121 galaxies which presenting an upturns of \ewha, \ewhda and/or a drop of \dindex\ with decreasing radius in their central regions. Such ``turnover'' features indicate recent/ongoing star formation in the central regions of these galaxies. We also examined global properties of the turnover galaxies, such as stellar mass and \nuvr\ colour, and structural properties, particularly the presence of a bar as well as environment.  We quantified the central turnover features and studied their correlation with global galaxy properties and bar properties.

Our conclusions can be summarised as follows.

\begin{enumerate}

\item  Central turnover features are found primarily in galaxies with stellar masses above $\sim$10$^{10}$ \msun\ and \nuvr\ colour less than 5.

\item  The majority of the turnover galaxies are barred, with a bar fraction of 89$\pm$3\% (108/121), much higher than the bar fraction of 34$\pm$4\% of a control sample of galaxies that are closely matched in stellar mass and colour. This reinforces previous findings that bar-driven instabilities lead to star formation enhancements in the central regions of galaxies.

\item For turnover galaxies with a bar, the radius of the central turnover region is found to positively correlate with the radius of the bar: \rt\ $=$ (0.35$\pm$0.02)\rbar+(0.61$\pm$0.13). This provides further evidence supporting a close link of central star formation enhancement and bars.
 
\item The central star formation enhancements can be quantified by comparing the observed and inward-extrapolated SFRs within turnover radii, and can be as high as $\approx$1 dex in our turnover galaxies.  These SFR enhancements show no obvious correlation with global properties such as stellar mass, colour, or location in the BPT diagram. 

\item There is no significant correlation of the turnover features with large-scale environment or the galaxy-galaxy interactions/mergers.

\end{enumerate}

\section*{Acknowledgements}

We thank the anonymous referee for comments that helped to improve the paper. 
This work was supported by the National Key R\&D Program of China (grant Nos. 2018YFA0404502) and the National Science Foundation of China (grant Nos. 11821303, 11973030, 11761131004, 11761141012, 11603075, U1831205 and 11703063).  LL thanks Sara Ellison for sharing the star formation enhancement catalog in their paper, thanks Yifei Luo for providing the $\Sigma_{1}$ calculation for MaNGA galaxies, thanks Ryan Chown for helpful comments on the manuscript.

This work makes use of data from SDSS-IV. Funding for SDSS-IV has been provided by the Alfred P. Sloan Foundation and Participating Institutions. Additional funding towards SDSS-IV has been provided by the US Department of Energy Office of Science. SDSS-IV acknowledges support and resources from the Centre for High-Performance Computing at the University of Utah. The SDSS web site is www.sdss.org.

SDSS-IV is managed by the Astrophysical Research Consortium for the Participating Institutions of the SDSS Collaboration including the Brazilian Participation Group, the Carnegie Institution for Science, Carnegie Mellon University, the Chilean Participation Group, the French Participation Group, Harvard–Smithsonian Center for Astrophysics, Instituto de Astrofsica de Canarias, The Johns Hopkins University, Kavli Institute for the Physics and Mathematics of the Universe (IPMU)/University of Tokyo, Lawrence Berkeley National Laboratory, Leibniz Institut fur Astrophysik Potsdam (AIP), Max-Planck-Institut fur Astronomie (MPIA Heidelberg), Max-Planck-Institut fur Astrophysik (MPA Garching), Max-Planck-Institut fur Extraterrestrische Physik (MPE), National Astronomical Observatory of China, New Mexico State University, New York University, University of Notre Dame, Observatario Nacional/MCTI, The Ohio State University, Pennsylvania State University, Shanghai Astronomical Observatory, United Kingdom Participation Group, Universidad Nacional Autonoma de Mexico, University of Arizona, University of Colorado Boulder, University of Oxford, University of Portsmouth, University of Utah, University of Virginia, University of Washington, University of Wisconsin, Vanderbilt University and Yale University.

This work makes use of HI-MaNGA data taken under proposal IDs: GBT16A-095, GBT17A-012 and GBT19A-127. The Green Bank Observatory is a facility of the National Science Foundation operated under cooperative agreement by Associated Universities, Inc.

This work makes use of the ALFALFA survey, based on observations made with the Arecibo Observatory. The Arecibo Observatory is operated by SRI International under a cooperative agreement with the National Science Foundation (AST-1100968), and in alliance with Ana G. M{\'e}ndez-Universidad Metropolitana, and the Universities Space Research Association. 

IRAF is distributed by the National Optical Astronomy Observatory, which is operated by the Association of Universities for Research in Astronomy (AURA) under a cooperative agreement with the National Science Foundation.




\bibliographystyle{mnras}
\bibliography{refs} 




\appendix

\section{The impact of spatial resolution on measurement of turnover radius}
\label{app:resolution}

\begin{figure}
	\begin{center}
		\includegraphics[width=0.4\textwidth]{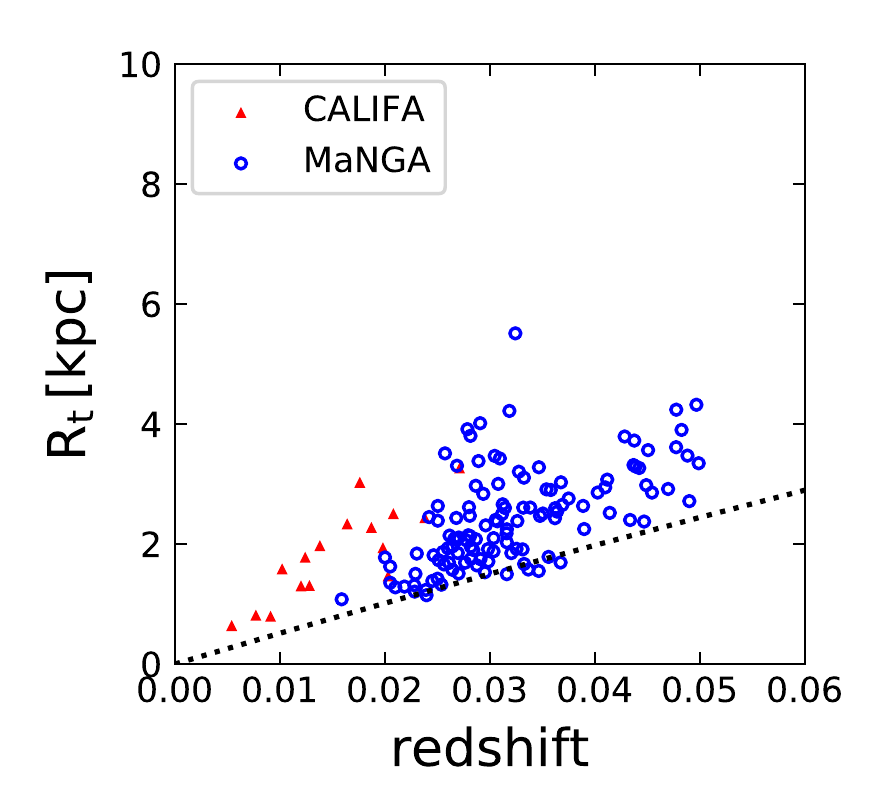}
		\includegraphics[width=0.4\textwidth]{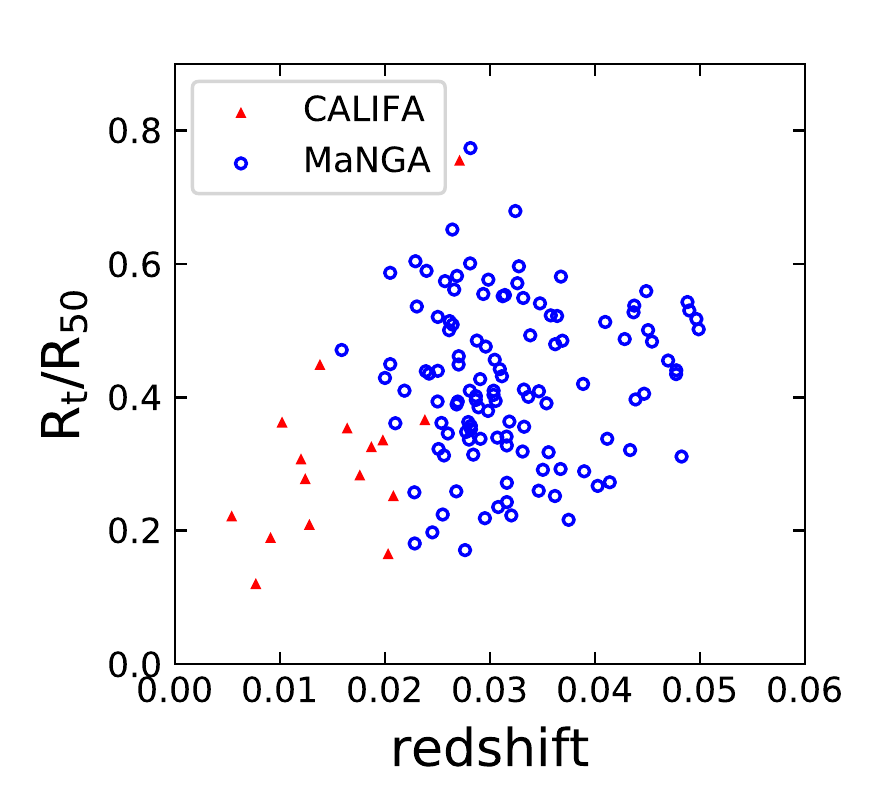}
	\end{center}
	\caption{Turnover radius \rt\ (upper) and the ratio of
		\rt/\re\ (lower) as a function of redshift. Blue circles
		are turnover galaxies from this work based on MaNGA, and red triangles
		are turnover galaxies from our previous work based on CALIFA.
		The dotted line indicates the physical scale at given redshift
		corresponding to 2\farcs5, the MaNGA spatial resolution.}
	\label{fig:appendix}
\end{figure}

One may worry about the limited spatial resolution of the MaNGA
data which may not be high enough for all the turnover radii to be
measured reliably. In Figure~\ref{fig:appendix} (upper panel) we
show the turnover radius \rt\ in unit of kpc  as a function of
redshift for the turnover galaxies in our sample (blue circles). The
black dotted line corresponds to 2.5$\arcsec$, i.e.  the spatial
resolution of MaNGA. As can be seen, all the turnover galaxies are
well above the dotted line, thus ensuring that the turnover features
are well resolved in our galaxies. On the other hand, however, we
see that the turnover radius is clearly correlated with the spatial
resolution, or equivalent the redshift.  This is not a real
correlation, but a result of the correlation of galaxy size (or
mass) with redshift, which is a combined effect of the fixed field
of view of MaNGA IFUs and the sample selection strategy.  The IFUs
of MaNGA have five different sizes, and the galaxy targets are
selected so as to have their angular size matched to the IFU, so
that the IFU covers out to 1.5\re\ or 2.5\re\ of each galaxy.  Due to
this selection, the average luminosity, stellar mass and size of the
MaNGA sample is an increasing function of redshift (see
Figure~\ref{fig:resolution}). This leads to the lack of turnover
galaxies with large \rt\ at lower redshifts as seen in
Figure~\ref{fig:appendix}, where  the physical size of the galaxies
and their central features (bulges, bars, turnovers, etc.) are
smaller. In the figure we also show the turnover galaxies from the
CALIFA survey \citep{Lin-17}, which has better resolution than MaNGA
and is limited to even lower redshifts. The same correlation is seen
between \rt\ and redshift. The CALIFA also uses IFU with a fixed FoV
of $\sim$1 arcmin, so it is also expected to see correlations of the
(physical) sizes of galaxies and their internal features/structures
with the redshift.

In the lower panel of Figure~\ref{fig:appendix}, 
we show the ratio of \rt\ relative to the effective radius
\re. We find no obvious correlation with the redshift, 
demonstrating that the correlation of \rt\ with redshift in the upper
panel is not real. In addition, we have done a analysis by
re-constructing the control sample by closely matching the
control sample with the turnover galaxy sample in redshift,
in addition to stellar mass and color. Our results and conclusions
remain unchanged. Therefore, we conclude that our results
are not affected by the spatial resolution of the data. 

\bsp	
\label{lastpage}
\end{document}